\documentclass[preprint,11pt,double-space,showpacs,preprintnumbers,showkeys,prd,nofootinbib,amsmath,amssymb]{revtex4}

\usepackage{psfrag,graphicx}
\usepackage{amssymb} 
\usepackage{amsfonts}
\usepackage{graphicx}

\begin{document}

\newcommand{\PIQ}{\ensuremath{\Pi(Q^{2})} }
\newcommand{\MD}{\ensuremath{\mathcal{D}} }
\newcommand{\MR}{\ensuremath{\mathcal{R}} }
\newcommand{\MK}{\ensuremath{\mathcal{K}} }
\newcommand{\MU}{\ensuremath{\mathcal{U}} }
\newcommand{\MC}{\ensuremath{\mathcal{C}} }
\newcommand{\omP}{\omega_{\Pi}}
\newcommand{\omD}{\omega_{\MD}}
\newcommand{\omK}{\omega_{\MK}}
\newcommand{\omU}{\omega_{\MU}}
\newcommand{\xih}{\hat{\xi}}
\newcommand{\QQ}{\ensuremath{Q^{2}}}
\newcommand{\MDlpt}{\ensuremath{\MD^{(L)}_{PT}} }
\newcommand{\non}{\nonumber}

\title{Infrared freezing of Euclidean QCD observables}

\author{Paul M. Brooks}
\email{p.m.brooks@durham.ac.uk}
\author{C. J. Maxwell}
\email{c.j.maxwell@durham.ac.uk}

\affiliation{Institute for Particle Physics Phenomenology, University of
  Durham, South Road, DH1 3LE, UK.}

\date{\today}

\begin{abstract}
We consider the leading one-chain term in a skeleton expansion
for QCD observables and show that for energies ${Q}^{2}>{\Lambda}^{2}$, where
${Q}^{2}={\Lambda}^{2}$ is the Landau pole of the coupling, the skeleton
expansion result is equivalent to the standard Borel integral representation, 
with ambiguities related to infrared (IR) renormalons. For ${Q^2}<{\Lambda}^{2}$ the
skeleton expansion result is equivalent to a previously proposed modified
Borel representation where the ambiguities are connected with ultraviolet
(UV) renormalons. We investigate the $Q^2$-dependence of the perturbative
corrections to the Adler-$D$ function, the GLS sum rule and the polarised and unpolarised Bjorken
sum rules. In all these cases the one-chain result changes sign in the
vicinity of ${Q}^{2}={\Lambda}^{2}$, and then exhibits freezing behaviour, vanishing at
${Q}^{2}={0}$. Finiteness at ${Q}^{2}={\Lambda}^{2}$ implies specific relations between the residues 
of IR and UV renormalon singularities in the Borel plane. These relations,
only one of which has previously been noted (though it remained unexplained), are shown to follow from the continuity of the 
characteristic function in the skeleton expansion. By considering the compensation
of non-perturbative and perturbative ambiguities we are led to a result for
the $Q^2$-dependence of these observables at all $Q^2$, in which there is a
single undetermined non-perturbative parameter, and which involves the skeleton
expansion characteristic function. The observables freeze to zero in the infrared.
We briefly consider the freezing behaviour of the Minkowskian $R_{{e}^{+}{e}^{-}}$ ratio.
\end{abstract}



\preprint{IPPP/06/28}
\preprint{DCPT/06/56}

\maketitle


\section{Introduction}
\setlength{\baselineskip}{18pt}
Thanks to asymptotic freedom, fixed-order QCD perturbation theory can potentially
provide accurate approximations to physical observables at suitably large energy
scales, $Q^2$. Such a perturbative description necessarily breaks down below the
Landau singularity at ${Q^2}={\Lambda}^{2}$, and the infrared behaviour unavoidably
involves non-perturbative effects. In fact non-perturbative information is needed
even to make sense of perturbation theory, since higher perturbative coefficients
exhibit factorial growth, and the perturbation series is not convergent. 
Using a Borel integral to represent the
resummed perturbation series, the Borel integral is ambiguous due to singularities
on the integration contour along the positive real semi-axis in the Borel plane,
so-called infrared (IR) renormalons. These ambiguities
are structurally the same as terms in the operator product expansion (OPE) in
powers of ${\Lambda}^{2}/{Q^2}$. OPE ambiguities and Borel representation
ambiguities can compensate each other, allowing the perturbative Borel and non-perturbative
OPE components to be separately well-defined once a regulation of the Borel integral,
such as principal value (PV), has been chosen \cite{r1}. For ${Q}^{2}<{\Lambda}^{2}$, however, the
Borel representation which is correlated with terms in the OPE breaks down. In a
recent paper Ref.[2], which focussed on the infrared freezing of the Minkowskian ${R}_{{e}^{+}{e}^{-}}$
ratio, it was suggested that below ${Q}^{2}={\Lambda}^{2}$ one should use a modified
Borel representation whose ambiguities come from singularities lying on the integration contour along the
negative real semi-axis, so-called ultraviolet (UV) renormalons. This Borel representation
has ambiguities which are structurally the same as a modified expansion in powers of
${Q}^{2}/{\Lambda}^{2}$, and once regulated both components can remain defined in
the infrared. This change of Borel representation has been claimed not to be   
physically motivated in Ref.\cite{r3}, where different conclusions about infrared
behaviour are reached. In this paper we shall show that if we postulate a QCD
skeleton expansion \cite{r4,r5}, then the leading one-chain term reproduces the standard
Borel representation for ${Q^2}>{\Lambda}^{2}$, and the proposed modified Borel
representation for ${Q}^{2}<{\Lambda}^{2}$. 

We consider the infrared behaviour
of the one-chain result for some Euclidean QCD observables. 
We shall concentrate on the Adler-$D$ function, the GLS sum rule and the polarised and unpolarised Bjorken sum rules \cite{r6,r7}. 
The skeleton expansion result automatically freezes to zero as $Q^2\rightarrow{0}$.
For the observables we consider, the freezing to zero occurs
after the Borel resummed perturbative corrections to the parton model result change sign in the vicinity
of ${Q}^{2}={\Lambda}^{2}$. Individual renormalon contributions to the Borel
integral diverge at ${Q}^{2}={\Lambda}^{2}$, but we find that when all of
the renormalons are summed over, one obtains a finite result. This finiteness requires
relations between the residues of infrared and ultraviolet renormalons. Only
one of these relations has
previously been noted \cite{r8}, and we show that they arise from
the continuity of the characteristic function in the skeleton expansion.
Considering the compensation of perturbative and OPE ambiguities alluded to above,
we are led to an expression for the $Q^2$-dependence of the observable written in
terms of the characteristic function, and containing a single undetermined non-perturbative
parameter. This result freezes to zero in the infrared. Existing discussions of infrared freezing
behaviour have largely focused on the Analytic Perturbation Theory (APT) approach \cite{r8A}. In this formalism
one expands observables in a basis of functions which have smooth infrared behaviour. For Euclidean
observables the unphysical Landau singularity in the coupling is cancelled by a power-like correction.
In contrast in our discussion finiteness and continuity emerge thanks to a subtle interplay
between UV and IR renormalons.

The plan of the paper is as follows. In Section 2 we shall introduce the QCD skeleton
expansion, and show that the one-chain leading term is equivalent to the
standard Borel representation for ${Q}^{2}>{\Lambda}^{2}$, and to the modified
representation for ${Q}^{2}<{\Lambda}^{2}$. We discuss what can be learnt about
the infrared freezing of observables. In Section 3 we describe the Borel plane
renormalon structure for our chosen Euclidean observables, and we show that finiteness
at ${Q^2}={\Lambda}^{2}$ only holds if there are cancellations between the residues
of IR and UV renormalons, the cancellations rely on a previously unknown relation
between the IR and UV residues. We write down a result for the $Q^2$ dependence of
the resummed observables in terms of Exponential Integral ($\textrm{Ei}$) functions, and plot the infrared
freezing behaviour to zero noted above. In Section 4 we consider
the skeleton expansion for the Adler $D$ function, and give an expression for the
characteristic function of the leading one-chain term. Making a power series expansion,
and changing variables, we explicitly obtain the Borel representations, and relate the
IR and UV renormalon residues to the power series coefficients of the characteristic function.
Continuity of the characteristic function 
is shown to underwrite
the relations between UV and IR renormalon residues noted above. In Section 5 we derive the
result for $Q^2$-dependence including non-perturbative effects mentioned above. 
In Section 6 we briefly
consider Minkowskian observables, specifically ${R}_{{e}^{+}{e}^{-}}$, and modify some of
the conclusions of Ref.\cite{r2} in the light of the criticisms of Ref.\cite{r3}. Section 7 contains a
discussion and our conclusions.

\section{QCD skeleton expansion and Borel representations}
\begin{figure*}
\includegraphics[width=0.85\textwidth]{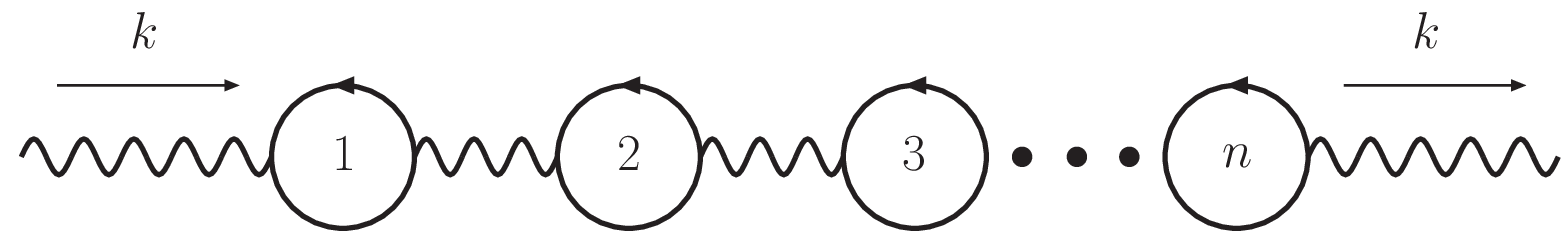}
\caption{A chain of fermion bubbles with momentum $k$ running through them.}
\label{F:fermionbubble}
\end{figure*}

Consider a generic Euclidean QCD observable ${\MD}(Q^2)$ having the perturbative
expansion
\begin{equation}
{\MD}_{PT}({Q}^{2})={a}({Q}^{2})+{\sum_{n>0}}{d}_{n}{a}^{n+1}({Q}^{2})\;.
\label{eq:1}\end{equation}
Here $a(Q^2)\equiv{\alpha}_{s}(Q^2)/{\pi}$ is the renormalised coupling. Throughout this
paper we will use the one-loop approximation for the coupling,
\begin{equation}
a(Q^{2})=\frac{2}{b\ln(Q^{2}/\Lambda^{2})}\;,
\label{eq:2}\end{equation}
where $b=(33-2{N}_{f})/6$ is the leading beta-function coefficient in SU($3$) QCD with $N_f$ active
quark flavours.
$Q^{2}\equiv-{q}^{2}>0$ is the single spacelike energy scale. As ${Q}^{2}\rightarrow{\infty}$ asymptotic
freedom ensures that ${\MD}(Q^2)\rightarrow{0}$. Our interest is in the infrared limit
${Q}^{2}\rightarrow{0}$, and the infrared behaviour of ${\MD}(Q^2)$. Specifically, is it
possible that freezing to a finite infrared limit ${\MD}(0)$ occurs? 
This is an intrinsically non-perturbative question which cannot be answered by perturbation theory
alone. One has in addition the non-perturbative contribution arising from the operator product
expansion (OPE),
\begin{equation}
{\MD}_{NP}({Q}^{2})={\sum_{n}}{\MC}_{n}{\left(\frac{{\Lambda}^{2}}{Q^2}\right)}^{n}\;.
\label{eq:3}\end{equation}
The freezing limit, if any, of ${\MD}(Q^2)={\MD}_{PT}(Q^2)+{\MD}_{NP}(Q^2)$,
depends on the behaviour of {\it both} components as ${Q}^{2}\rightarrow{0}$. 
Perturbative freezing will not arise from fixed-order perturbation theory, one needs an
all-orders resummation of Eq.(\ref{eq:1}). Unfortunately our exact information about the higher-order
coefficients is limited, at best, to calculations of $d_1$ and $d_2$, higher-orders are unknown. All-orders
information is only available in the large-${N}_{f}$ limit where one expands each $d_n$ as
\begin{equation}
{d}_{n}={d}_{n}^{[n]}{N}_{f}^{n}+{d}_{n}^{[n-1]}{N}_{f}^{n-1}+\ldots+{d}_{n}^{[0]}\;.
\label{eq:4}\end{equation}
The leading large-${N}_{f}$ coefficient ${d}_{n}^{[n]}$ can be computed exactly to all-orders
since it arises from a restricted set of Feynman diagrams in which a chain of $n$
fermion bubbles (shown in Fig. \ref{F:fermionbubble})
is inserted in a basic skeleton diagram \cite{r9,r10}. In principle one can consider more
than one chain and construct a QED skeleton expansion \cite{r11}. In QCD one can replace ${N}_{f}$ by $(33/2-3b)$
, and obtain an expansion in powers of $b$,
\begin{equation}
{d}_{n}={d}_{n}^{(n)}{b}^{n}+{d}_{n}^{(n-1)}{b}^{n-1}+\ldots+{d}_{n}^{(0)}\;.
\label{eq:5}\end{equation}
The leading-$b$ term ${d}_{n}^{(L)}\equiv{d}_{n}^{(n)}{b^n}$ can then be used
to approximate $d_n$ \cite{r8,r8a,r8b} and an all-orders resummation of these terms performed to obtain ${\MD}_{PT}^{(L)}(Q^2)$. Use of the
one-loop form of the coupling in Eq.(\ref{eq:2}) ensures that this resummed result is RS-independent.

The leading term of the skeleton expansion arises from integrating over the momentum $k$ flowing through
the chain of bubbles \cite{r4,r5,r12}.
\begin{equation}
\MDlpt(Q^2)=\int_{0}^{\infty}{dt}\;{\omega}(t)a({e}^{C}{t}{Q}^{2})\;.
\label{eq:6}\end{equation}
Here $t\equiv{k}^{2}/{Q}^{2}$, and $\omega(t)$ is the so-called characteristic function
of the observable. The constant $C$ depends on the subtraction procedure used to
renormalise the bubble. Standard ${\overline{MS}}$ subtraction corresponds to $C=-5/3$.
>From now on we shall assume $C=0$ which corresponds to the so-called $V$-scheme,
${\overline{MS}}$ subtraction with renormalization scale ${\mu}^{2}={e}^{-5/3}{Q}^{2}$.
$\Lambda$ in Eq.(\ref{eq:2}) will refer to that in the $V$-scheme. The characteristic function
satisfies the normalization condition 
\begin{equation}
\int_{0}^{\infty}{dt}\;{\omega}(t)=1\;,
\label{eq:7}\end{equation}
which ensures the leading $a(Q^2)$ coefficient of unity assumed in Eq.(\ref{eq:1}).
The form of $\omega(t)$ changes at
$t=1$, and the range of integration splits into an IR and a UV part
\begin{equation}
\MDlpt(Q^2)=\int_{0}^{1}{dt}\;{\omega}_{IR}(t)a(t{Q}^{2})+\int_{1}^{\infty}{dt}\;{\omega}_{UV}(t)a(t{Q}^{2})\;,
\label{eq:8}\end{equation}
the IR part corresponding to ${k}^{2}<{Q}^{2}$, and the UV part to ${k}^{2}>{Q}^{2}$. By making a change of variable
one can transform the leading skeleton term into a Borel representation. For ${Q}^{2}>{\Lambda}^{2}$ one has the
standard Borel representation (we shall explicitly write down the required changes of variable in Sec. 4),
\begin{equation}
\MDlpt({Q}^{2})={\int_{0}^{\infty}}{dz}\,{e}^{-z/a({Q}^{2})}B[\MDlpt](z)\;.
\label{eq:9}\end{equation}
Here $B[{\MD}](z)$ is the Borel transform, defined by,
\begin{equation}
B[{\MD}_{PT}^{(L)}](z)={\sum_{n=0}^{\infty}}\frac{{z}^{n}{d}^{(L)}_{n}}{n!}\;.
\label{eq:10}\end{equation}
$B[{\MD}_{PT}^{(L)}](z)$ contains singularities along the real $z$-axis. In the large-$b$ approximation
these are single and double poles at positions $z={z}_{n}$ and $z=-{z}_{n}$, with ${z_n}\equiv{2n/b}$, $n=1,2,3\ldots$. The singularities
on the positive real semi-axis are referred to as {\it infrared} renormalons, ${IR}_{n}$, and those on the negative
real semi-axis as {\it ultraviolet} renormalons, ${UV}_{n}$. The ${IR}_{n}$ renormalons cause the Borel
representation to be ambiguous since they lie on the integration contour along the positive
real $z$-axis. The difference between routing the contour above or below the singularity yields an ambiguity
\begin{equation}
\Delta{\MD}_{PT}^{(L)}\sim{\left(\frac{{\Lambda}^{2}}{Q^2}\right)}^{n}\;,
\label{eq:11}\end{equation}
which has the same form as a term in the OPE in Eq.(\ref{eq:3}), so that OPE ambiguities associated with the
${({\Lambda}^{2}/{Q}^{2})}^{n}$ OPE term in ${\MD}_{NP}(Q^2)$ can potentially cancel against the ${IR}_{n}$ renormalon
ambiguity allowing each component separately to be well defined \cite{r1}. In practice we shall choose to
take a Principal Value (PV) definition of the integral. The IR part of the $t$ integration in Eq.(\ref{eq:8})
produces the IR renormalon part of the Borel representation, and 
needs to be PV regulated. The second UV component produces the UV renormalons and does not require
regulation. As we shall see in the next section the standard Borel representation of Eq.(\ref{eq:9}) for
Euclidean quantities diverges like $\ln{a}(Q^2)$ at ${Q^{2}}={\Lambda}^{2}$ for each {\it individual} ${IR}_{n}$ or
${UV}_{n}$ renormalon contribution. When the full set is resummed, however, the $\ln{a}$ divergence
is cancelled and a finite result is found. We shall explore this further in
Sections 3 and 4. 

For ${Q}^{2}<{\Lambda}^{2}$,
${a}(Q^2)<0$, and the representation of Eq.(\ref{eq:9}) is invalid. The key point is that the change of variable
from $t$ to $z$ is proportional to $a(Q^2)$, and so if ${a}(Q^2)$ changes sign the limits of
integration in $z$ change sign, yielding the modified Borel representation 
\begin{equation}
\MDlpt({Q}^{2})=\int_{0}^{-\infty}{dz}\,{e}^{-z/a(Q^2)}B[\MDlpt](z)\;.
\label{eq:12}\end{equation}
This is the modified Borel representation proposed in Ref.\cite{r2} where it was motivated
as a standard Borel representation corresponding to an expansion in $|a(Q^2)|=-a(Q^2)$, since by
changing variables one can write Eq.(\ref{eq:12}) as
\begin{equation}
\MDlpt({Q}^{2})=-\int_{0}^{\infty}{dz}\,{e}^{-z/|a(Q^2)|}B[\MDlpt](-z)\;.
\label{eq:13}\end{equation}
So we see that the one-chain skeleton contribution of Eq.(\ref{eq:6}) is equivalent to the standard
Borel representation of Eq.(\ref{eq:9}) for ${Q^2}>{\Lambda}^{2}$, and to the modified representation
of Eq.(\ref{eq:12}) for ${Q^2}<{\Lambda}^{2}$. Note that when we substitute
Eq.~(\ref{eq:10}) into the Borel representation of Eq.~(\ref{eq:13}), then it reproduces the correct form
of the perturbative expansion in Eq.~(\ref{eq:1}), for negative $a$.
The modified Borel representation now has a contour of integration along the {\it negative} real semi-axis,
and so it is rendered ambiguous by the ultraviolet ${UV}_{n}$ renormalon singularities. Correspondingly 
the IR component of Eq.(\ref{eq:8}) is now well-defined and it is now the UV component which requires regulation.
The ambiguity from routing the contour is now
\begin{equation}
\Delta{\MD}_{PT}^{(L)}(Q^2)\sim{\left(\frac{Q^2}{\Lambda^2}\right)}^{n}\;.
\label{eq:14}\end{equation}

It was suggested in Ref.\cite{r2} that the usual OPE of Eq.(\ref{eq:3}) breaks down for ${Q^2}<{\Lambda}^{2}$,
as does the associated PT Borel representation of Eq.(\ref{eq:9}), and should be recast and replaced by
a modified expansion in powers of $Q^2/{\Lambda}^{2}$,
\begin{equation}
{\MD}_{NP}({Q}^{2})=\sum_{n}{\tilde{\MC}}_{n}{\left(\frac{{Q}^{2}}{{\Lambda}^{2}}\right)}^{n}\;.
\label{eq:15}\end{equation}
The $n\rm{th}$ term in this expansion has then structurally the same form as the ambiguity associated with the 
${UV}_{n}$ renormalon contribution. It was further suggested in Ref.\cite{r2} that a ${\tilde{\MC}}_{0}$ term
independent of $Q^2$ could arise from rearrangement of the standard OPE. This was motivated by a simple
toy example. In fact in the one-chain approximation no such term arises and both PT and NP components freeze
to zero. The terms in Eq.(\ref{eq:15}) are then in one-to-one correspondence with the ${UV}_{n}$ renormalon ambiguities.
From its definition, the QCD skeleton expansion implies ${\MD}_{PT}(0)=0$ in the $Q^2\rightarrow{0}$
limit. For the one-chain term in Eq.(\ref{eq:6}) this simply follows because as ${Q}^{2}\rightarrow{0}$
the integrand vanishes everywhere in the range of integration, since $a(tQ^2)\rightarrow{0}$ for any given $t$.
Higher multiple chain terms will contain products of the form $a({t}_{1}{Q^2})a({t}_{2}{Q^2})\ldots$ in the
integrand and will similarly vanish. This then implies that in the infrared limit ${\MD}_{NP}(Q^2)$
behaves as
\begin{equation}
{\MD}_{NP}(Q^2)\approx k{\left(\frac{Q^2}{{\Lambda}^{2}}\right)}^{n_0}\;,
\label{eq:16}\end{equation}
where ${UV}_{n_0}$ is the UV renormalon singularity nearest to the origin in the Borel plane.
We should note that the
modified Borel representation, its infrared behaviour and its connection with UV renormalons, has
also been discussed in Ref.\cite{r13}. In Appendix B of that paper the infrared freezing of the Adler function $D(Q^2)$ was
discussed and it was concluded that from general arguments of non-perturbative spontaneous chiral symmetry breaking in the limit
of a large number of colours, $N_c$, one expected that as $Q^2\rightarrow{0}$, $D(Q^2)\rightarrow{0}$ like
\begin{equation}
D(Q^2)\sim\frac{Q^2}{M^2}\;,
\label{eq:17}\end{equation}
where $M$ is the mass of a one-meson state, these states remaining massive in the chiral limit.
A similar result is obtained in Ref.\cite{r15a}.
Since ${UV}_{1}$ is
the singularity nearest the origin for the Adler function,
${n}_{0}=1$, and the freezing expectation is indeed consistent with Eq.(\ref{eq:16}). Notice
that strictly the leading behaviour as $Q^2\rightarrow{0}$ is the logarithmic freezing to zero
of $a(Q^2)$ contributed by the PT component. It is the non-perturbative effects which reflect
the UV renormalon structure.  

\section{$Q^2$-dependence of the Euclidean observables}
\begin{figure*}
\begin{center}
\begin{tabular}{lr}
\includegraphics[angle=270,width=0.45\textwidth]{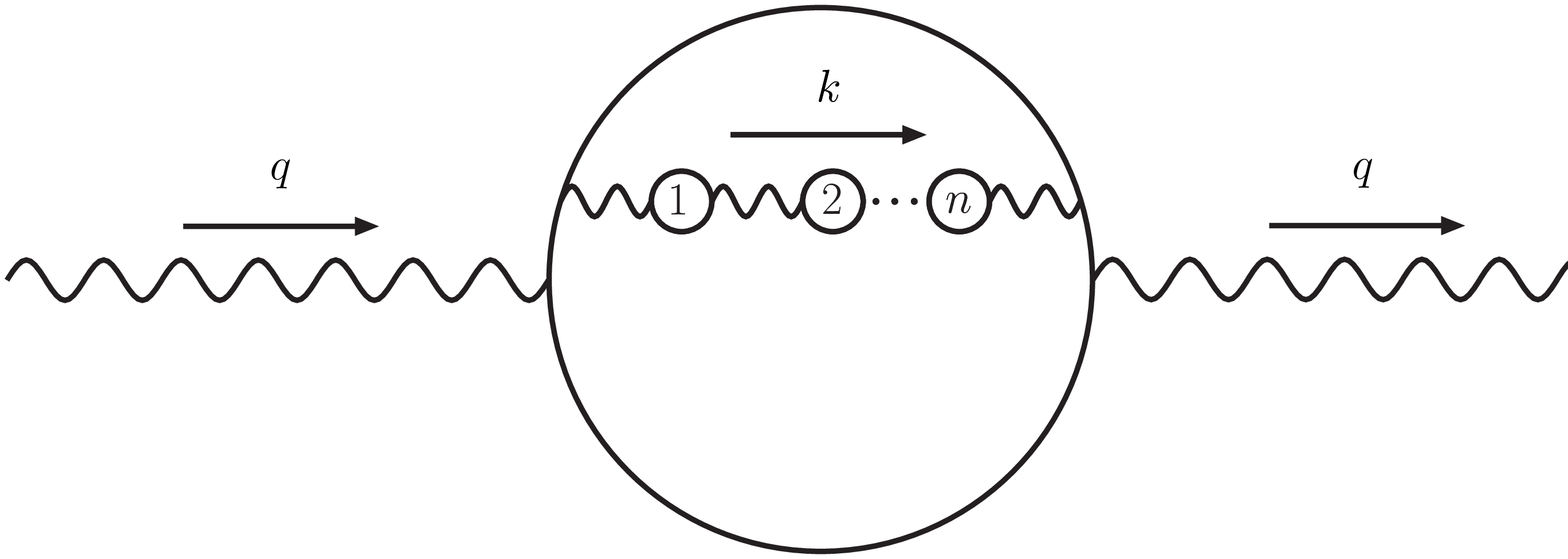}\hspace{.05\textwidth}&\hspace{.05\textwidth}
\includegraphics[angle=270,width=0.45\textwidth]{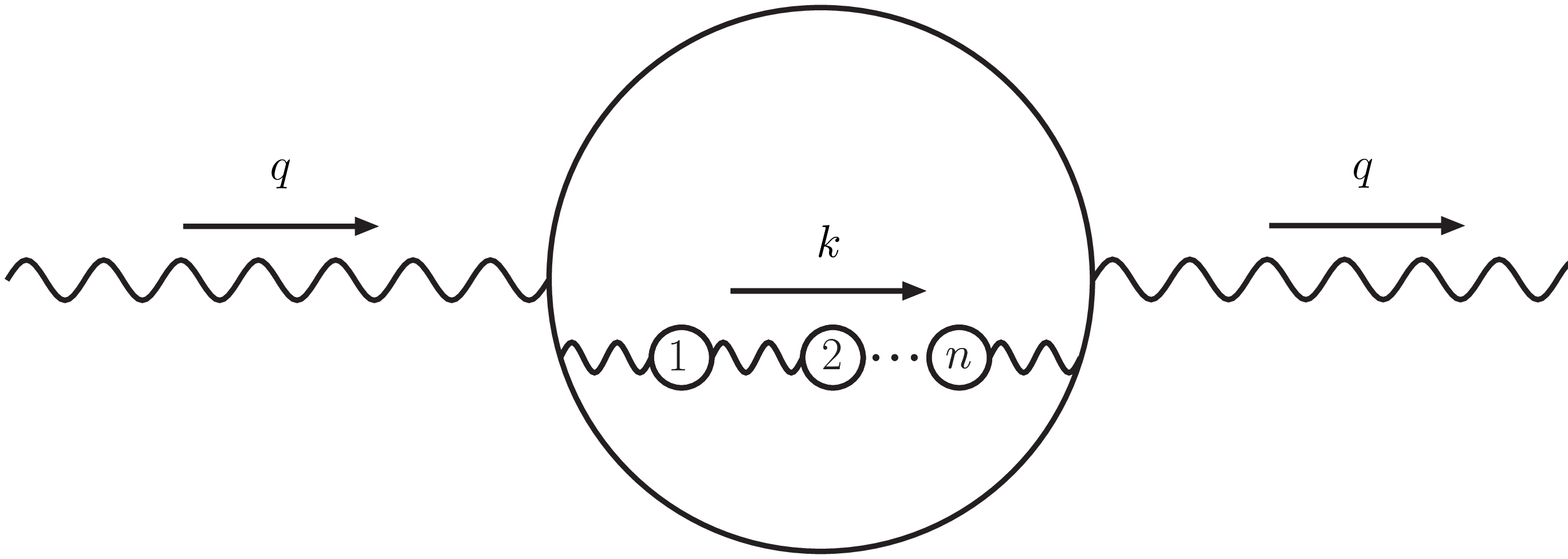}\vspace{-2.5cm}\\
\multicolumn{2}{c}{
\includegraphics[angle=270,width=0.45\textwidth]{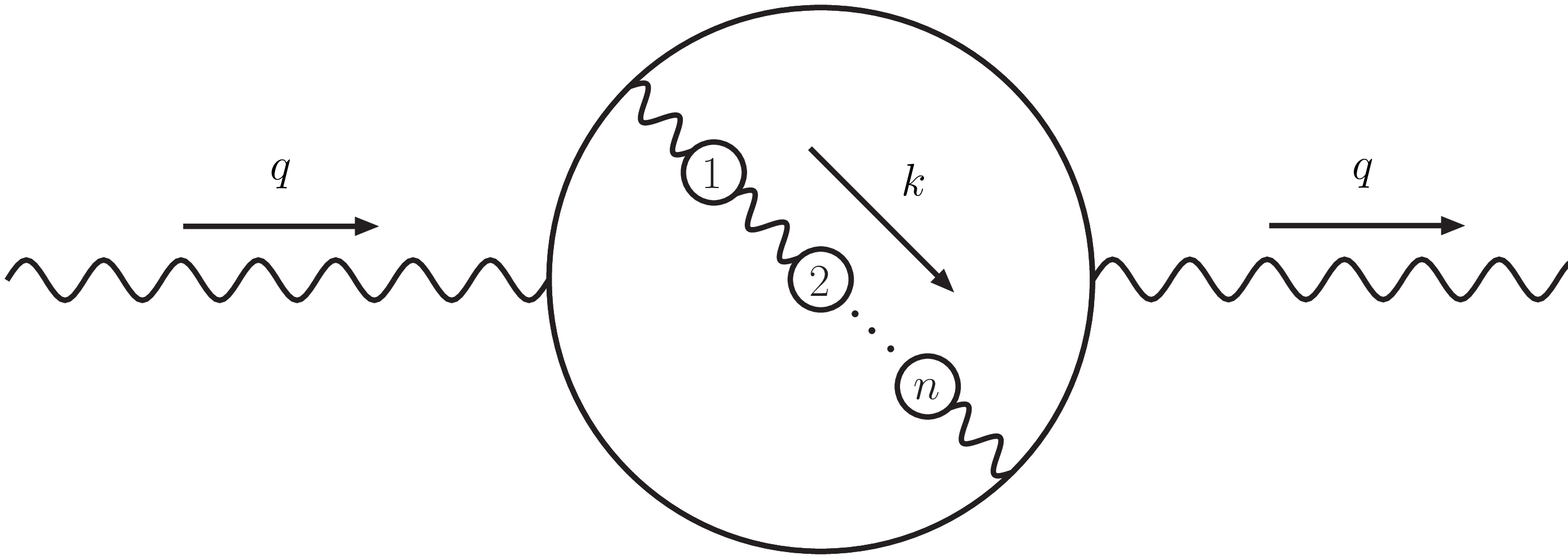}}\vspace{-2cm}\\
\end{tabular}
\end{center}
\caption{Leading large-$N_{f}$ contributions to the vacuum polarisation function at $n$th order in perturbation theory.}
\label{F:vacuumpolarizationNf}
\end{figure*}

\begin{figure*}
\begin{tabular}{lcr}
\includegraphics[width=0.3\textwidth]{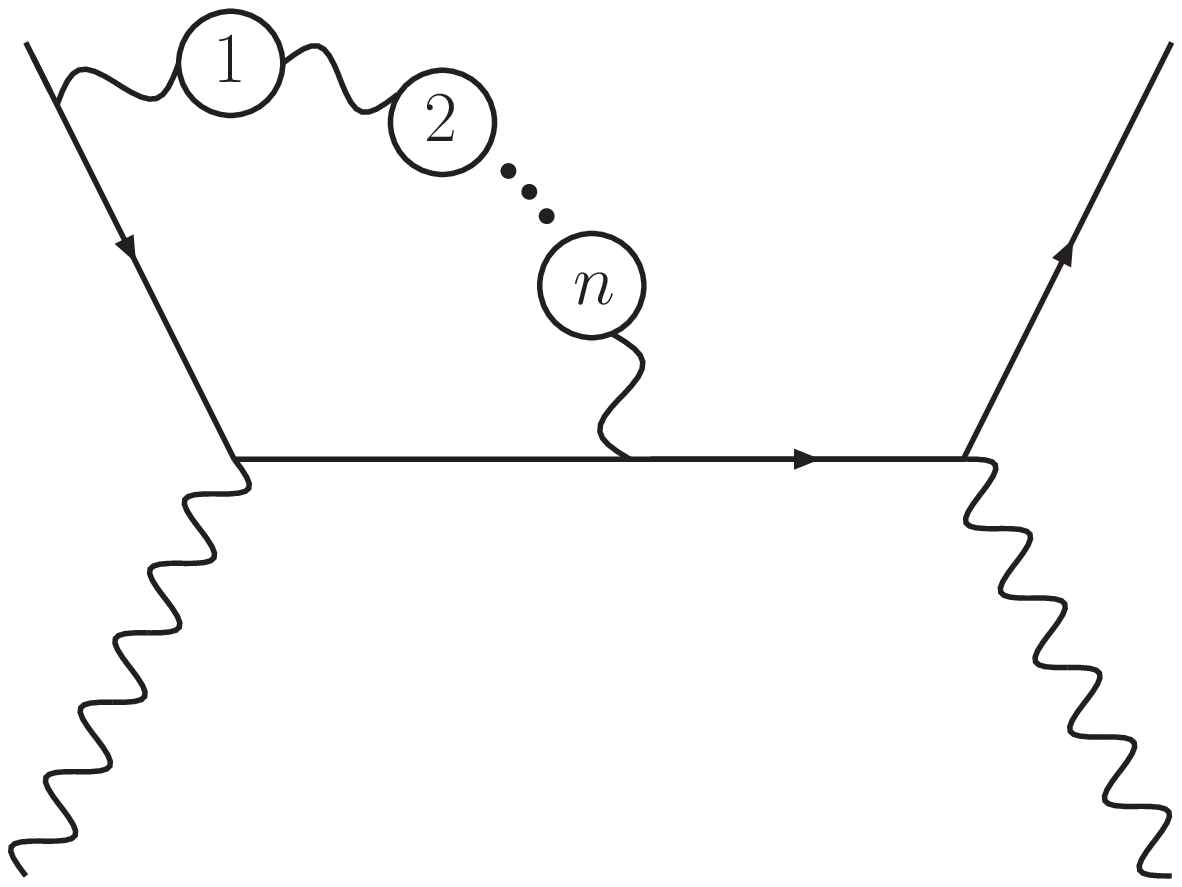}&\includegraphics[width=0.3\textwidth]{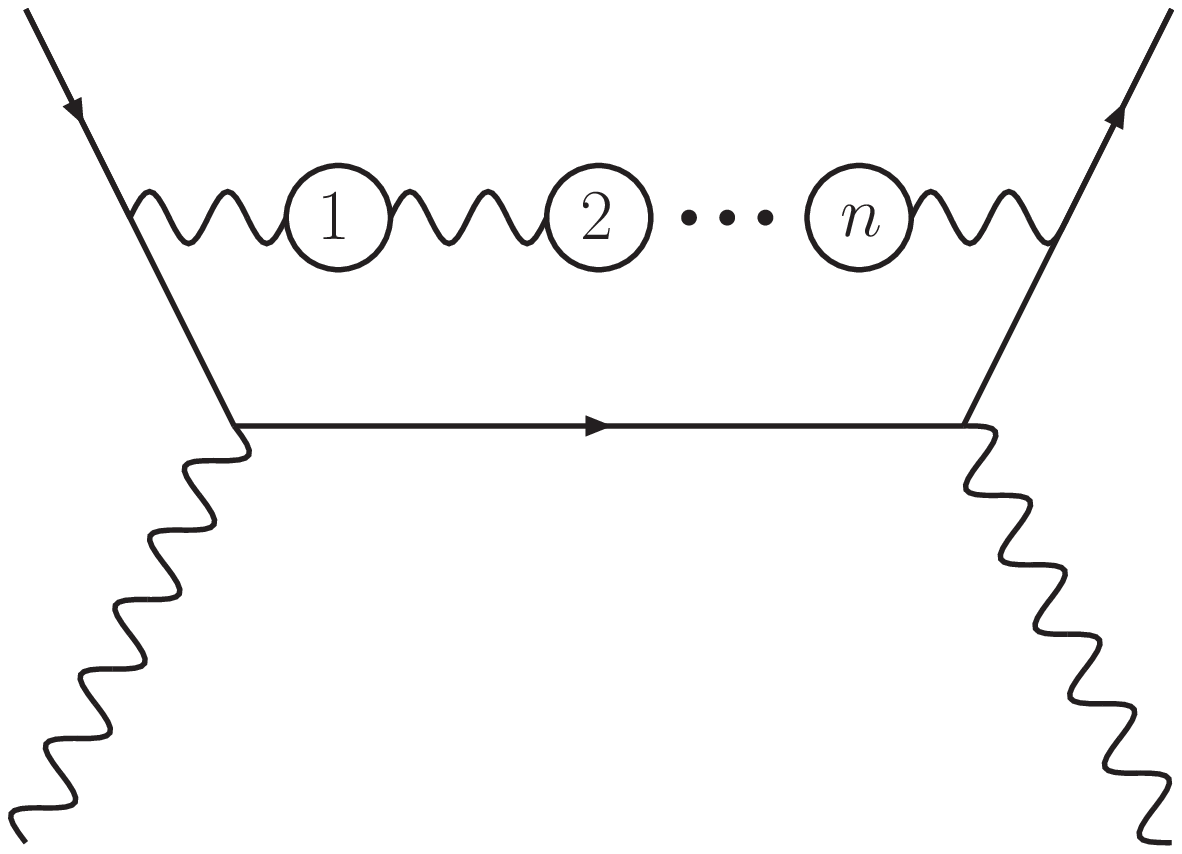}&
{\includegraphics[width=0.3\textwidth]{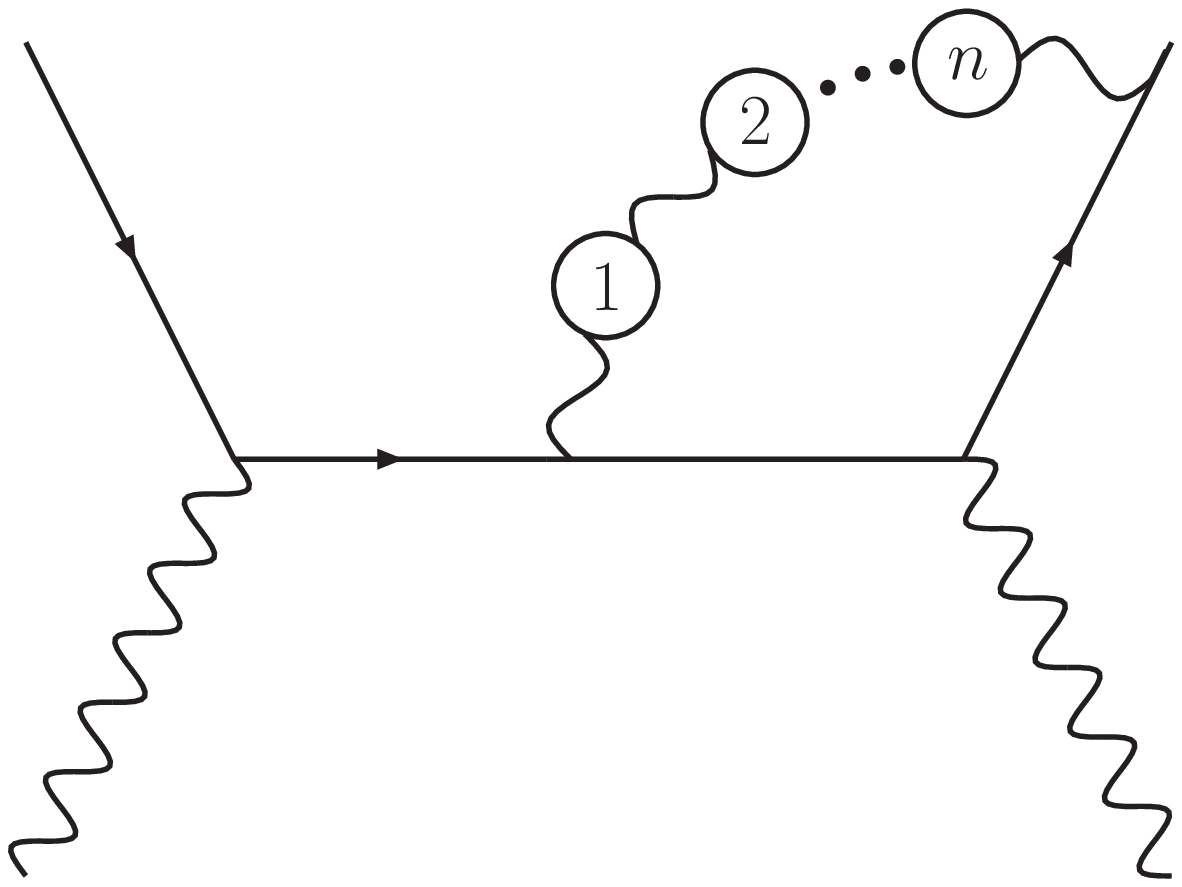}}\\
\end{tabular}
\caption{Leading large-$N_{f}$ contributions to the DIS sum rules of
  Eqs.~(\ref{eq:22}) - (\ref{eq:24}) at $n$th order in perturbation theory.}
\label{F:sumrulesNf}
\end{figure*}

We begin by defining the three Euclidean observables we shall consider.
The QCD vacuum polarization function, $\Pi(Q^{2})$, is the
correlator of two vector currents in the Euclidean region,
\begin{eqnarray}
(q_{\mu}q_{\nu}-g_{\mu\nu}q^{2})\Pi(Q^{2})=16\pi^{2}i\int
d^{4}x\;e^{iq.x}\langle0|T[j_{\mu}(x)j_{\nu}(0)]|0\rangle,
\label{eq:19}\end{eqnarray}
The leading-$N_{f}$ component of $\Pi(Q^{2})$ can be calculated from the
diagrams in Fig. \ref{F:vacuumpolarizationNf}.
The Adler function, $D(Q^{2})$, is then defined via the logarithmic derivative
of $\Pi(Q^{2})$
\begin{eqnarray}
D(Q^{2})&=&-\frac{3}{4}Q^{2}\frac{d}{dQ^{2}}\Pi(Q^{2}).\label{D}
\label{eq:20}\end{eqnarray}
This can be split into  the parton model result and QCD
corrections, $\MD(Q^2)$,
\begin{eqnarray}
D(Q^{2})=N_{c}\sum_{f}Q_{f}^{2}\Bigg{(}1+\frac{3}{4}C_{F}\MD(Q^2)\bigg{)},
\label{eq:21}\end{eqnarray}
where $N_{c}$ is the number of colours, $C_{F}=\frac{(N_{c}^{2}-1)}{2N_{c}}$, and $Q_f$ is the charge of quark flavour $f$.
Here ${\MD}(Q^2)={\MD}_{PT}(Q^2)+{\MD}_{NP}(Q^2)$, with the two components defined as in Eqs.~(\ref{eq:1}) and (\ref{eq:3}).
The polarised Bjorken (pBj) \cite{r14} and GLS \cite{r15} sum rules are
defined as
\begin{eqnarray}
K_{pBj}&\equiv&\int_{0}^{1}g_{1}^{ep-en}(x,Q^{2})dx\non \\
&=&\frac{1}{3}\Bigg{|}\frac{g_{A}}{g_{V}}\Bigg{|}\Bigg{(}1-\frac{3}{4}C_{F}\MK(Q^2)\Bigg{)}\;,
\label{eq:22}\end{eqnarray}
\begin{eqnarray}
K_{GLS}&\equiv&\frac{1}{6}\int_{0}^{1}F_{3}^{\bar{\nu}p+\nu p}
(x,Q^{2})dx\non \\
&=&\Bigg{(}1-\frac{3}{4}C_{F}\MK(Q^2)\Bigg{)}\label{Kg}\;.
\label{eq:23}\end{eqnarray}
$\MK(Q^2)$ being the QCD corrections to the parton model
result, again split into PT and NP components as for ${\MD}(Q^2)$. We have neglected contributions due to
``light-by-light'' diagrams -- which when omitted render the
perturbative corrections to $K_{GLS}$ and $K_{pBj}$ identical.
Finally, the unpolarised Bjorken sum rule (uBj) \cite{r16} is defined as
\begin{eqnarray}
U_{uBj}&\equiv&\int_{0}^{1}F_{1}^{\bar{\nu}p-\nu p}(x,Q^{2})dx \non \\
&=&\Bigg{(}1-\frac{1}{2}C_{F}\MU(Q^2)\Bigg{)}\;. \label{Ub}
\label{eq:24}\end{eqnarray}
The QCD corrections to the parton model result are again split into PT and NP components.
The leading-$N_{f}$ contributions to these three sum rules can be calculated
from the diagrams in Fig. \ref{F:sumrulesNf}. These large-$N_f$ results can be used to compute
leading-$b$ all-orders resummations for these observables, ${\MD}^{(L)}_{PT}(Q^2)$, ${\MK}^{(L)}_{PT}(Q^2)$ and ${\MU}^{(L)}_{PT}(Q^2)$,
as described in Section 2.
\\
The Borel transform of $\MDlpt(Q^2)$ 
is well-known and can be found
in Ref.\cite{r8},
\begin{eqnarray}
B[\MDlpt](z)
&=&\sum_{n=1}^{\infty}\frac{A_{0}(n)-A_{1}(n)z_{n}}
{\Big{(}1+\frac{z}{z_{n}}\Big{)}^{2}}+\frac{A_{1}(n)z_{n}}
{\Big{(}1+\frac{z}{z_{n}}\Big{)}}\non\\
&+&\sum_{n=1}^{\infty}\frac{B_{0}(n)+B_{1}(n)z_{n}}
{\Big{(}1-\frac{z}{z_{n}}\Big{)}^{2}}-\frac{B_{1}(n)z_{n}}
{\Big{(}1-\frac{z}{z_{n}}\Big{)}}\label{BorelD}\;.
\label{eq:25}\end{eqnarray}
Here
\begin{eqnarray}
A_{0}(n)&=&\frac{8}{3}\frac{(-1)^{n+1}(3n^{2}+6n+2)}{n^{2}(n+1)^{2}(n+2)^{2}},\qquad\qquad
A_{1}(n)\;\;=\;\;\frac{8}{3}\frac{b(-1)^{n+1}(n+\frac{3}{2})}{n^{2}(n+1)^{2}(n+2)^{2}}\non\\[10pt]
&&B_{0}(1)=0,\quad B_{0}(2)=1,\quad B_{0}(n)=-A_{0}(-n)\quad
n\geq3\non \\
&&B_{1}(1)=0,\quad B_{1}(2)=-\frac{b}{4},\quad B_{1}(n)=-A_{1}(-n)\quad
n\geq3
\label{eq:26}\end{eqnarray}
These definitions coincide with Ref.\cite{r8},
except for $B_{1}(2)=-\frac{b}{4}$.  The
purpose of the slight change of definition 
is to make more explicit the single and double pole structure.
The Borel transforms of ${\MK}^{(L)}_{PT}(Q^2)$ and ${\MU}^{(L)}_{PT}(Q^2)$ can be found in Refs.\cite{r6,r7}, respectively.
They have a much simpler structure than that of the Adler-$D$ function since
they arise from insertion of the chain of bubbles into a tree-level diagram, rather
than into a quark loop, as shown in Fig. \ref{F:sumrulesNf}. There are
only a finite number of single poles and no double poles. Consequently we
can write out their Borel transforms explicitly
\begin{eqnarray}
B[{\MK}^{(L)}_{PT}](z)=\frac{4/9}{\Big{(}1+\frac{z}{z_{1}}\Big{)}}-\frac{1/18}{\Big{(}1+\frac{z}{z_{2}}\Big{)}}+\frac{8/9}{\Big{(}1-\frac{z}{z_{1}}\Big{)}}-\frac{5/18}{\Big{(}1-
\frac{z}{z_{2}}\Big{)}}\;.
\label{BorelK}
\label{eq:27}\end{eqnarray}
and
\begin{eqnarray}
B[{\MU}^{(L)}_{PT}](z)=\frac{1/6}{\Big{(}1+\frac{z}{z_{2}}\Big{)}}+\frac{4/3}{\Big{(}1-\frac{z}{z_{1}}\Big{)}}-\frac{1/2}{\Big{(}1-\frac{z}{z_{2}}\Big{)}}\;.
\label{BorelU}
\label{eq:28}\end{eqnarray}
As noted in Refs.\cite{r7,r8} the leading-$b$ approximations for the NLO and NNLO coefficients for these observables are
in reasonable agreement with the known exact coefficients.

We can now evaluate the Borel integral of Eq.(\ref{eq:9}) to obtain $\MDlpt(Q^2)$, ${\MK}^{(L)}_{PT}(Q^2)$ and ${\MU}^{(L)}_{PT}(Q^2)$. 
Using the integrals
\begin{eqnarray}
\int_{0}^{\infty}{dz}\frac{e^{-z/a}}{(1+z/z_{n})}=-z_{n}e^{z_{n}/a}\textrm{Ei}(-z_{n}/a)\label{BI1}\;,\label{eq:29}\\
\int_{0}^{\infty}{dz}\frac{e^{-z/a}}{(1+z/z_{n})^{2}}=z_{n}\Bigg{[}1+\frac{z_{n}}{a}e^{-z_{n}/a}\textrm{Ei}(z_{n}/a)\Bigg{]}\label{BI2}\;,
\label{eq:30}\end{eqnarray}
the following resummed expressions are obtained,
\begin{eqnarray}
\MDlpt(Q^2)&=&\sum_{n=1}^{\infty}z_{n}\left\{e^{z_{n}/a(Q^2)}\textrm{Ei}\left(\!\!-\!\frac{z_{n}}{a(Q^2)}\right)\left[\frac{z_{n}}{a(Q^2)}\left(A_{0}(n)-z_{l}A_{1}(n)\right)-z_{n}A_{1}(n)
\right]\right.\non\\
&&\hspace{3cm}+\left(A_{0}(n)-z_{n}A_{1}(n)\right)\Bigg{\}}\non\\
&+&\sum_{n=1}^{\infty}z_{n}\Bigg{\{}e^{-z_{n}/a(Q^2)}\textrm{Ei}\left(\frac{z_{n}}{a(Q^2)}\right)\Bigg{[}\frac{z_{n}}{a(Q^2)}\left(B_{0}(n)+z_{l}B_{1}(n)\right)-z_{n}B_{1}(n)\Bigg{]}\non\\
&&\hspace{3cm}-
\left(B_{0}(n)+z_{n}B_{1}(n)\right)\Bigg{\}}\;,\label{MD}
\label{eq:31}\end{eqnarray}
\begin{eqnarray}
{\MK}^{(L)}_{PT}(Q^2)=\frac{1}{9b} \Bigg{[}-8e^{{z_1}/a(Q^2)}\textrm{Ei}\left(-\frac{z_{1}}{a(Q^2)}\right)+2e^{z_{2}/a(Q^2)}\textrm{Ei}\left(-\frac{z_{2}}{a(Q^2)}\right)&&\non\\
+16e^{-z_{1}/a(Q^2)}\textrm{Ei}\left(\frac{z_{1}}{a(Q^2)}\right)-10e^{-z_{2}/a(Q^2)}\textrm{Ei}\left(\frac{z_{2}}{a(Q^2)}\right)\Bigg{]}&&\!\!\!,
\label{eq:32}\end{eqnarray}
\begin{eqnarray}
{\MU}^{(L)}_{PT}(Q^2)&=&\frac{1}{3b}\Bigg{[}8e^{-z_{1}/a(Q^2)}\textrm{Ei}\left(\frac{z_{1}}{a(Q^2)}\right)-6e^{-z_{2}/a(Q^2)}\textrm{Ei}\left(\frac{z_{2}}{a(Q^2)}\right)
\non \\
&-&2e^{z_{2}/a(Q^2)}\textrm{Ei}\left(-\frac{z_{2}}{a(Q^2)}\right)\Bigg{]}\;.
\label{eq:33}\end{eqnarray}
Where Ei$(x)$ is the exponential integral function defined (for $x<0$) as
\begin{eqnarray}
\textrm{Ei}(x)\equiv-\int_{-x}^{\infty}dt\frac{e^{-t}}{t}\;,
\label{eq:34}\end{eqnarray}
and for $x>0$ by taking the PV of the integral.
It has the expansion
\begin{eqnarray}
\textrm{Ei}(x)=\ln|x| +\gamma_{E}+\mathcal{O}(x)\label{EI exp}\;,
\label{eq:35}\end{eqnarray}
for small $x$, where $\gamma_{E}=0.57721\ldots$ is the Euler constant.\\

A crucial point is that the above expressions for the $Q^2$-dependence apply at {\it all} values
of $Q^2$. For ${Q^2}<{\Lambda}^{2}$ the modified Borel representation, written as an ordinary
Borel representation for an expansion in powers of $|a|$, as in Eq.(\ref{eq:13}), corresponds to changing $a(Q^2)\rightarrow-a(Q^2)$
, ${z}_{n}\rightarrow-{z}_{n}$, and adding an overall minus sign in
Eqs.(\ref{eq:31})-(\ref{eq:33}). One can easily
see that these equations are invariant under these changes. In Eq.(\ref{eq:31}) one needs to change ${A}_{1}\rightarrow-{A}_{1}$ and ${B}_{1}\rightarrow-{B}_{1}$, 
since they contain a hidden $z_n$ factor in their definitions, also in
Eqs.(\ref{eq:32}) and (\ref{eq:33}), the prefactor proportional
to $1/b$ also needs to change sign since it has been factorised from ${z}_{1},{z}_{2}$. 
The $\textrm{Ei}(z_n/a(Q^2))$ functions
exhibit a logarithmic divergence as their argument goes to zero, and so it would appear that one does not obtain
a finite result at $Q^2={\Lambda}^{2}$. Using Eq.(\ref{eq:35}), one has,
\begin{eqnarray}
\textrm{Ei}\Bigg{[}\frac{2n}{ba(Q^2)}\Bigg{]}&=&\textrm{Ei}\Bigg{[}n\log(Q^{2}/\Lambda^{2})\Bigg{]} \non \\
&\simeq&\textrm{Ei}\Bigg{[}n\Bigg{(}\frac{Q^{2}}{\Lambda^{2}}-1\Bigg{)}\Bigg{]} \non \\
&\simeq&\gamma_{E}+\ln\Bigg{[}n\Bigg{(}\frac{Q^{2}}{\Lambda^{2}}-1\Bigg{)}\Bigg{]}\label{apr}\;,
\label{eq:36}\end{eqnarray}
for ${\Lambda}^{2}\approx{Q}^{2}$.
Note that the only terms in Eqs.(\ref{eq:31})-(\ref{eq:33}) which could
possibly contribute to  the divergence are $e^{\pm
z_{n}/a}\textrm{Ei}(\mp z_{n}/a)$ terms and, as can be seen from
Eqs.(\ref{eq:29}) and (\ref{eq:30}),
these are generated
exclusively by the single pole terms in the Borel transform. The double
pole terms only generate finite contributions at
$Q^{2}=\Lambda^{2}$.

\begin{figure*}
\begin{center}
\begin{tabular}{lr}
\hspace{.15cm}
\includegraphics[angle=270,width=0.45\textwidth]{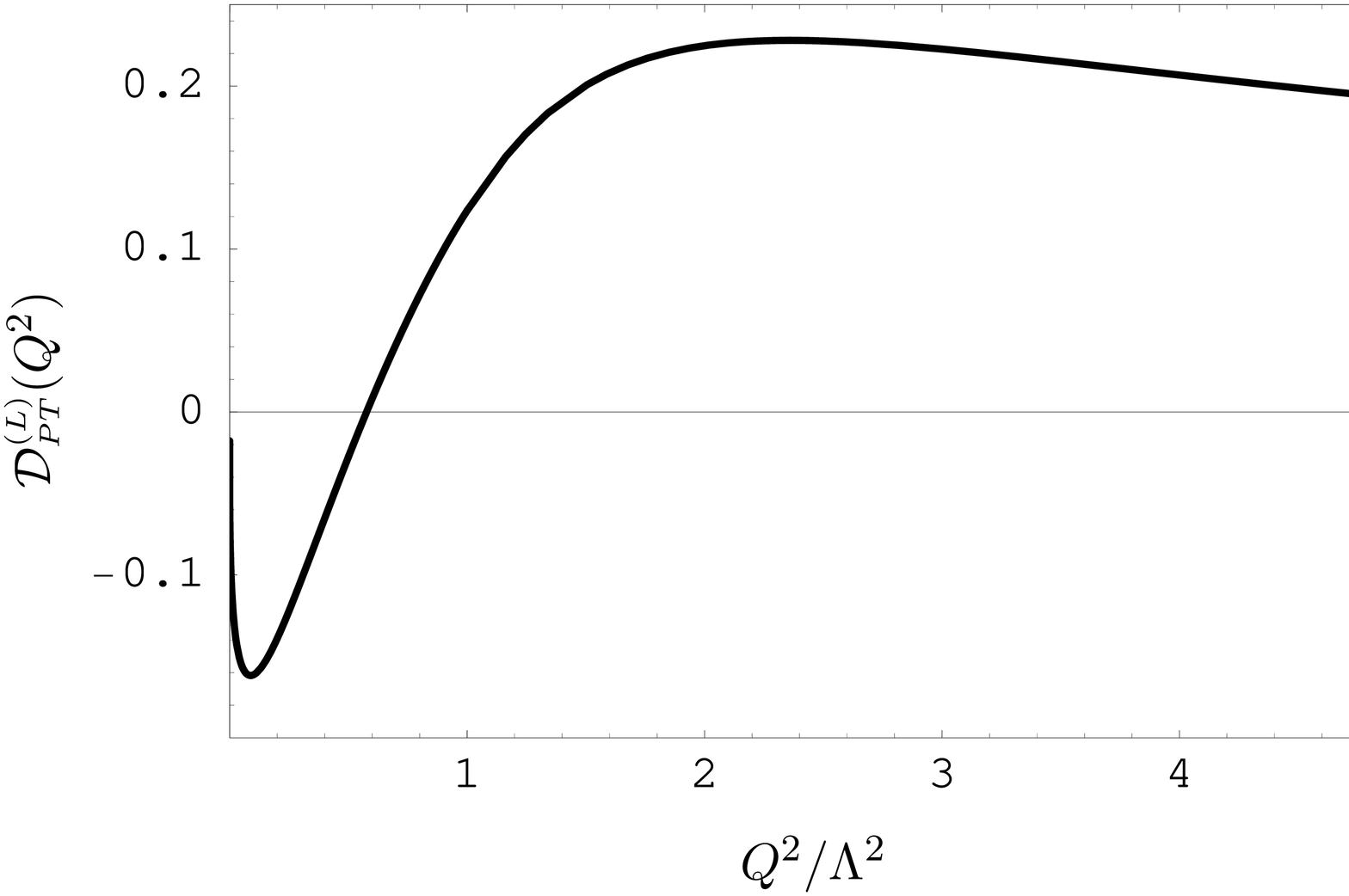}\hspace{.045\textwidth}&\hspace{.045\textwidth}
\includegraphics[angle=270,width=0.45\textwidth]{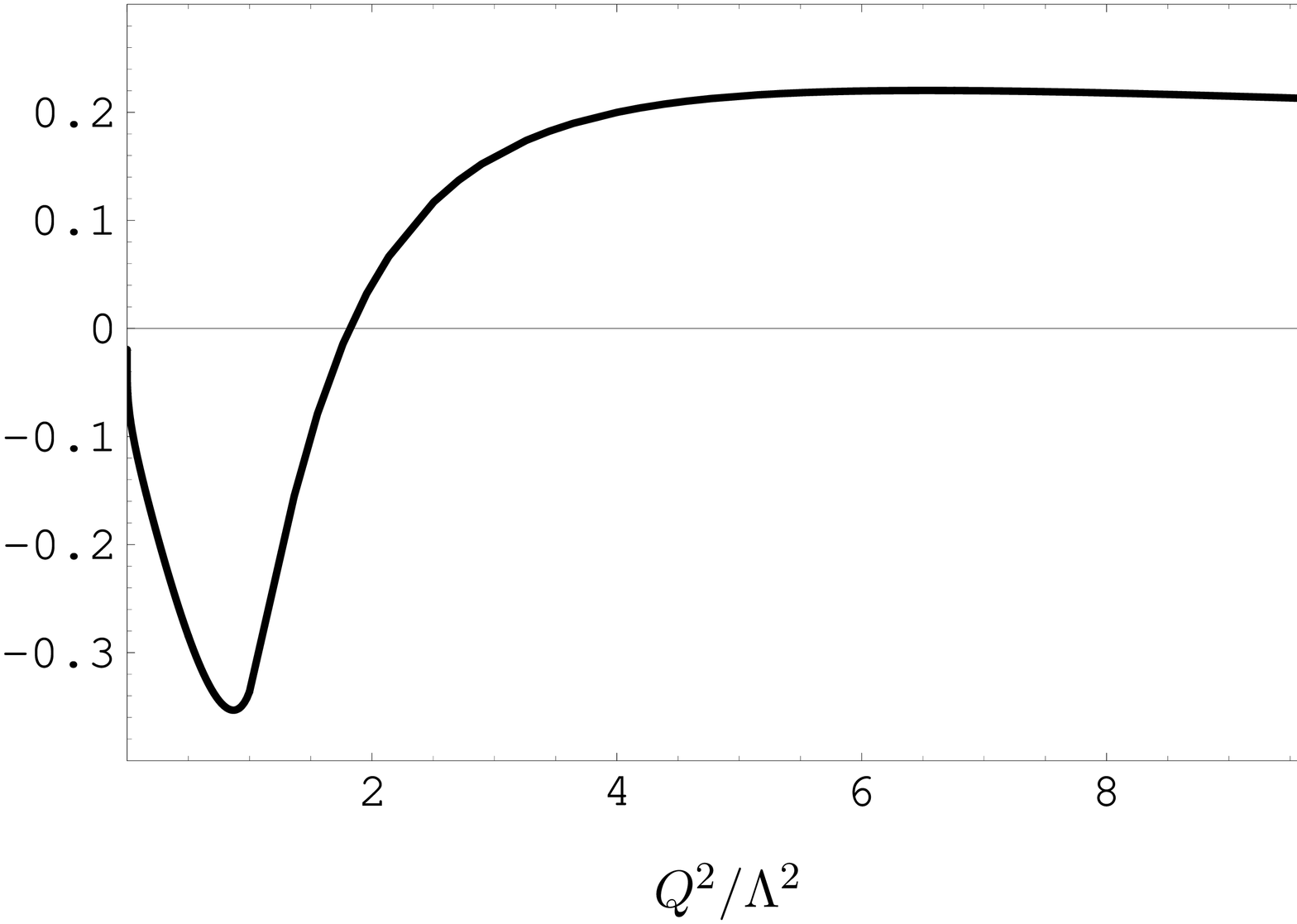}\\
\multicolumn{2}{c}{
\includegraphics[angle=270,width=0.45\textwidth]{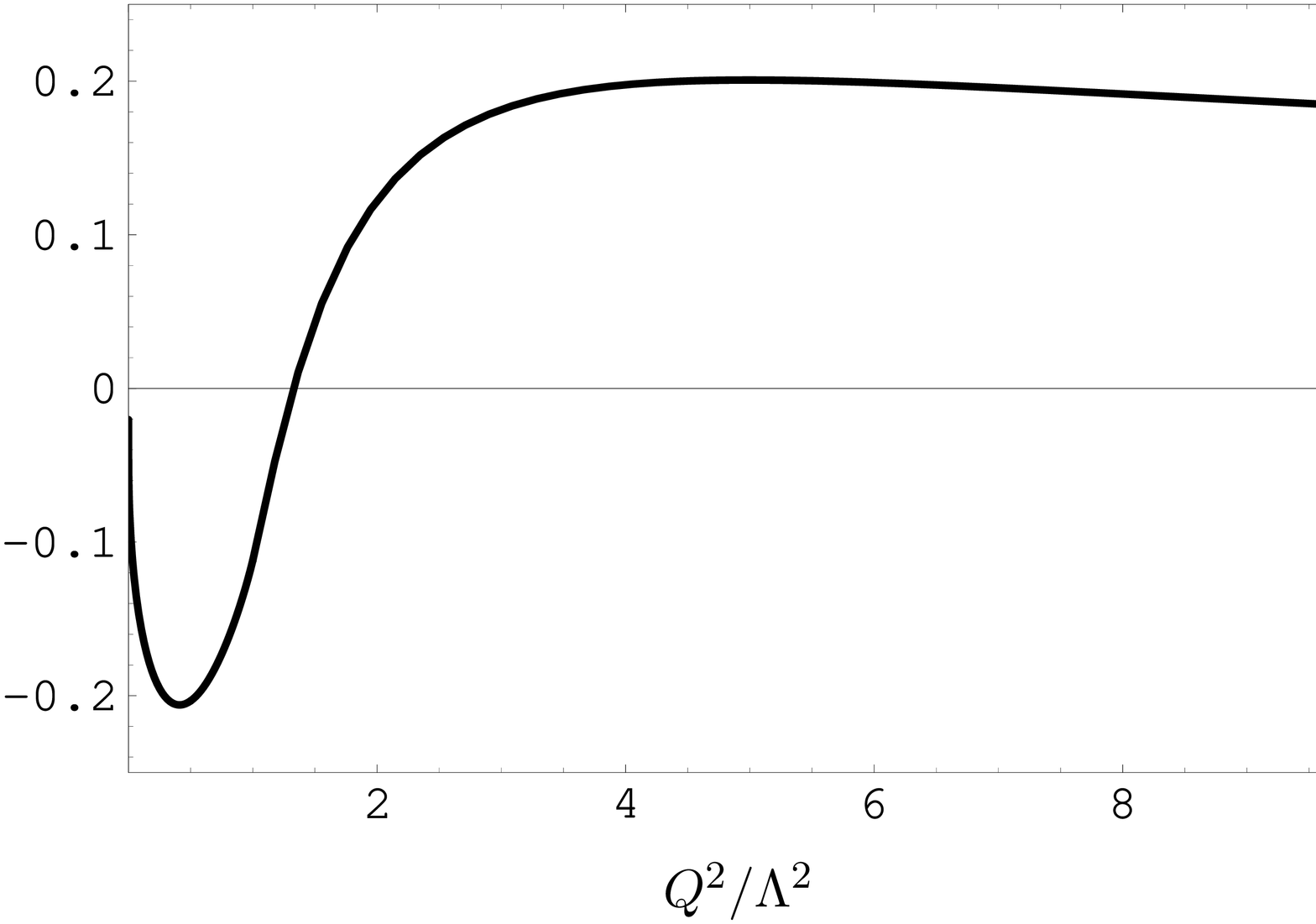}}
\end{tabular}
\end{center}
\caption{$Q^2$-dependence of the {\it perturbative} corrections to the observables
  in Eqs.(\ref{eq:21})-(\ref{eq:24}), resummed to
  all orders in the leading-$b$ approximation.}
\label{F:DUK}
\end{figure*}

Using Eq.(\ref{eq:36}) we obtain the $Q^2\rightarrow \Lambda^{2}$ limit of
$\MDlpt(Q^{2})$
\begin{eqnarray}
\MDlpt(Q^{2})&=&-\sum_{n=1}^{\infty}z^{2}_{n}[A_{1}(n)+B_{1}(n)]\ln\Bigg{[}n\Bigg{(}\frac{Q^{2}}{\Lambda^{2}}-1\Bigg{)}\Bigg{]}\label{D
at QL}\label{eq:37}\\
&+&\sum_{n=1}^{\infty}[z_{n}^{2}(1+\gamma_{E})(-A_{1}(n)-B_{1}(n))+z_{n}(A_{0}(n)-B_{0}(n))]+\mathcal{O}\Bigg{(}\frac{Q^{2}}{\Lambda^{2}}-1\Bigg{)}\non
\;.
\end{eqnarray}
So the coefficient of the divergent log term in $\MDlpt(Q^{2})$ is,
\begin{eqnarray}
-\sum_{n=1}^{\infty}z^{2}_{n}[A_{1}(n)+B_{1}(n)]\;,
\label{eq:38}\end{eqnarray}
and for ${\MK}^{(L)}_{PT}(Q^{2})$ and ${\MU}^{(L)}_{PT}(Q^{2})$ the
equivalent coefficients are
$(-8+2+16-10=0)$ and $(8-6-2=0)$, respectively. Cancellation clearly
occurs in the cases of ${\MK}^{(L)}_{PT}(Q^{2})$ and ${\MU}^{(L)}_{PT}(Q^{2})$ and in the
case of $\MDlpt(Q^{2})$ the previously unnoticed relation
\begin{eqnarray}
z_{n+3}^{2}B_{1}(n+3)=-z_{n}^{2}A_{1}(n)\;,
\label{n+3 relation}
\label{eq:39}\end{eqnarray}
ensures that $\MDlpt({\Lambda}^{2})$ is finite,
\begin{eqnarray}
\sum_{n=1}^{\infty}z^{2}_{n}[A_{1}(n)+B_{1}(n)]\label{a1+b1 sum}=0\;.
\label{eq:40}\end{eqnarray}
A similar relation
\begin{equation}
{A}_{0}(n)=-{B}_{0}(n+2)\;,
\label{eq:41}\end{equation}
was noted in \cite{r8}.
We shall show in the next section that the relations of Eqs.(\ref{eq:39}) and
(\ref{eq:41}) are underwritten
by the continuity of the skeleton expansion characteristic function
${\omega}_{\Pi}(t)$ and its first derivative at $t=1$.
The form of the perturbative corrections, $\MDlpt(Q^{2})$, ${\MK}^{(L)}_{PT}(Q^{2})$ and ${\MU}^{(L)}_{PT}(Q^{2})$, are shown  in
Fig. \ref{F:DUK}.

Although we have shown that when summed to infinity Eq.(\ref{eq:31}) is
finite at $Q^{2}=\Lambda^{2}$, we obviously can only plot the
expression including a finite number of terms in the $n$ sum. The expression can remain finite, however, if we sum the UV renormalons to finite $n=N$ and the IR
renormalons to $n=N+3$. In this case the relation of Eq.(\ref{eq:39})
will ensure that the divergent terms cancel.
We took $N=50$ and assumed ${N}_{f}=0$ quark flavours, avoiding the need to match at
quark flavour thresholds, since we are only interested here in the form of the freezing
behaviour, not in a phenomenological analysis.

The plots in Fig. \ref{F:DUK} demonstrate two important points about the
Euclidean quantities we are considering. Firstly the finite
behaviour at $Q^{2}=\Lambda^{2}$, and secondly that the Borel resummed perturbative corrections to the
parton model result change sign
just below or above this point. For $\MD$ these corrections become
negative but crucially the full observable $D(Q^2)$ remains positive at all values of $Q^2$. 
They then freeze to zero 
as noted in Sec.~2.\\

The relation of Eq.(\ref{eq:39}) simplifies the expression for the
finite part of Eq.(\ref{eq:37}), it becomes
\begin{eqnarray}
\MDlpt(Q^{2}={\Lambda}^{2})&=&\sum_{n=1}^{\infty}z_{n}[A_{0}(n)-B_{0}(n)]-\sum_{n=1}^{\infty}z^{2}_{n}[A_{1}(n)+B_{1}(n)]\ln
n\non\\
&\approx&0.123625\;.
\label{eq:42}\end{eqnarray}
The values ${\MK}^{(L)}_{PT}(\QQ ={\Lambda}^{2})$ and ${\MU}^{(L)}_{PT}(\QQ
={\Lambda}^{2})$ are given by a formula identical to Eq.(\ref{eq:42}), but using
values of $A_{0,1}(n)$ and $B_{0,1}(n)$ appropriate to $\MK$ and $\MU$.
Although we have not given these values explicitly, they are of a much simpler form than in the case of $\MD$, and
they can easily be deduced
by comparing Eqs.(\ref{eq:27}) and (\ref{eq:28}) with Eq.(\ref{eq:25}). From this we obtain,
\begin{eqnarray}
\MK_{PT}^{(L)}(\QQ ={\Lambda}^{2})=-\frac{8}{9b}\ln2\;,\non \\
\MU_{PT}^{(L)}(\QQ ={\Lambda}^{2})=-\frac{8}{3b}\ln2\;.
\label{eq:43}\end{eqnarray}

\section{Skeleton expansion and Borel representations for the Adler function}

\begin{figure*}
\begin{tabular}{lcr}
\includegraphics[width=0.3\textwidth]{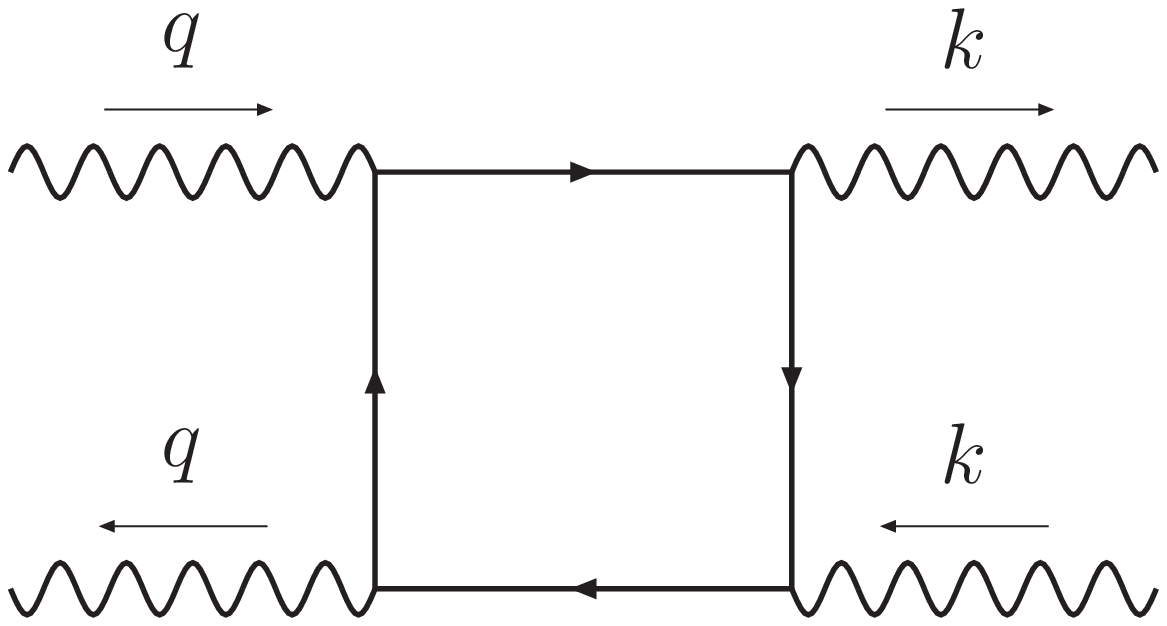}&\includegraphics[width=0.3\textwidth]{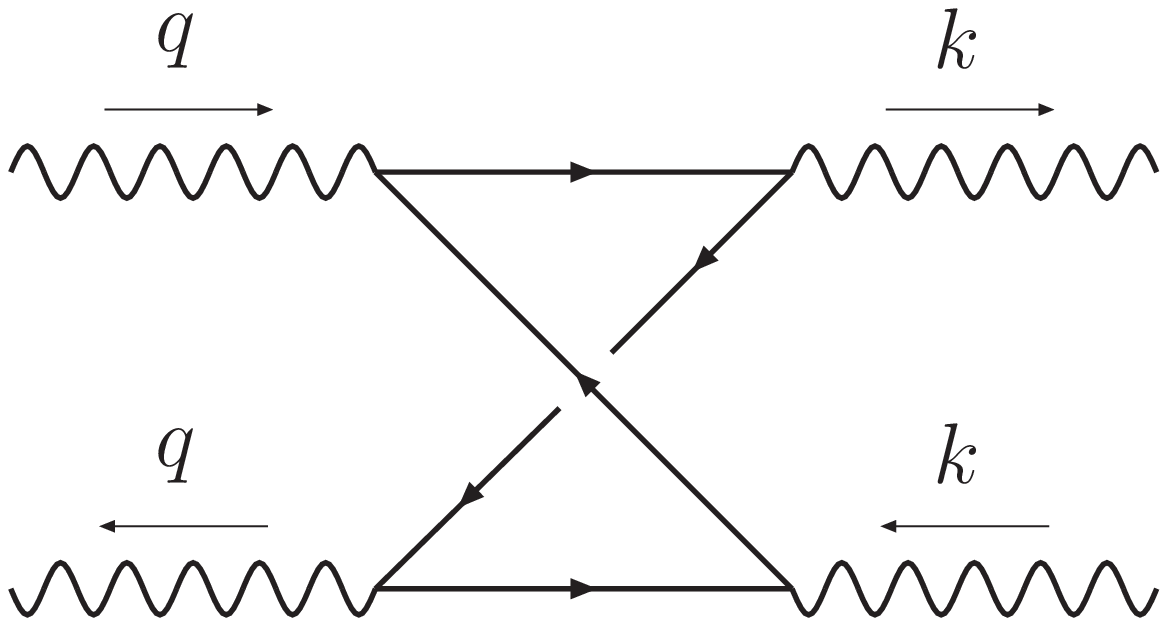}&{\includegraphics[width=0.3\textwidth]{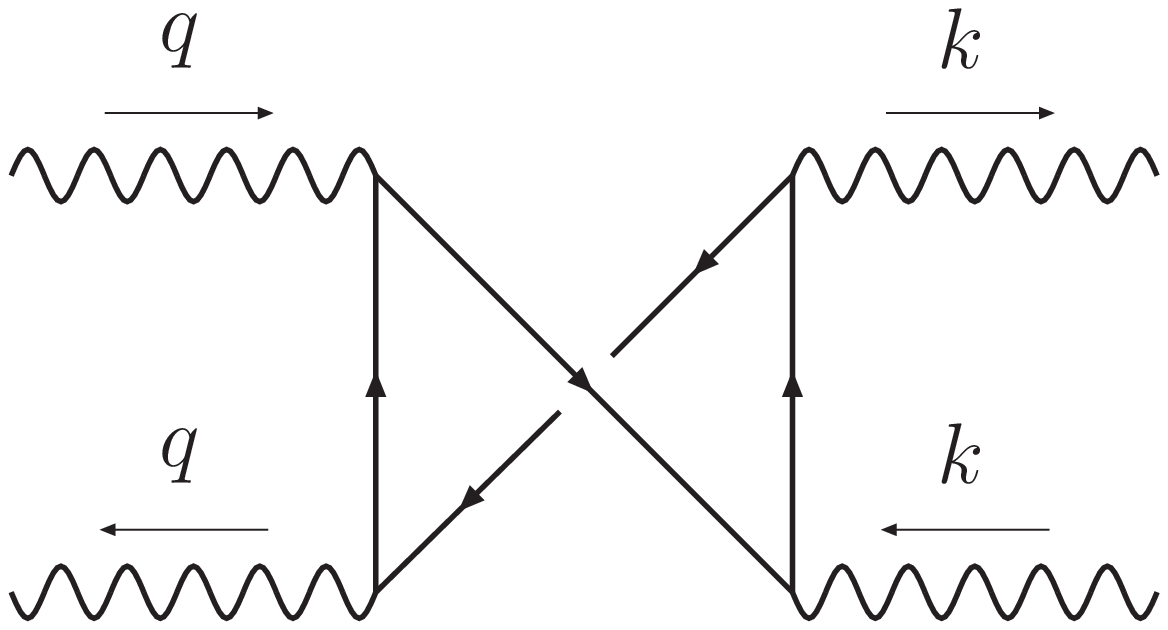}}\\
\end{tabular}
\caption{Light-by-light scattering diagrams, used to calculate $\omP$.}
\label{F:Kernel}
\end{figure*}

We begin with the one-chain skeleton expansion result for the vacuum polarization
function $\Pi(Q^2)$ defined in Eq.(\ref{eq:19}), 
\begin{eqnarray}
\Pi(Q^{2})=\int_{0}^{\infty}{dt}\;\omP(t)a(t Q^{2})\;,\label{Pi
spec}
\label{eq:44}\end{eqnarray}
where the characteristic function $\omP(t)$ is given by
\begin{eqnarray}
\omP(t)=-\frac{4}{3}
\left\{
    \begin{array}{ll}
    t\,\Xi(t)                                    &\hskip10mm     t\leq1\quad\leftrightarrow\quad\textrm{IR} \\
                                                &  \label{eq:45}\label{conf}\\
    \frac{1}{t}\,\Xi\Big{(}\frac{1}{t}\Big{)}     &\hskip10mm     t\geq1\quad\leftrightarrow\quad\textrm{UV} \\
    \end{array}
\right.
\end{eqnarray}
It can be obtained from the classic QED work of Ref.\cite{r17} by simply including
appropriate colour factors.\footnote{The origin of the minus sign in Eq.(\ref{eq:45}) is the
difference between the definitions of $\Pi$ given in Eq.~(\ref{eq:19}) and
Ref.\cite{r17}.} In this language it is related to the Bethe-Salpeter
kernel for the scattering of light-by-light, and is the first term in a well-defined
QED skeleton expansion \cite{r11}. The diagrams relevant to the kernel are
shown in Fig. \ref{F:Kernel}.
It is easy to see how, by connecting the ends of the fermion bubble chain in
Fig. \ref{F:fermionbubble} to the momentum $k$ external propagators in
Fig. \ref{F:Kernel}, one can reproduce the topology of the diagrams in
Fig. \ref{F:vacuumpolarizationNf}. The existence of the QCD skeleton expansion is more
problematic \cite{r12}. $\Xi(t)$ is given by \cite{r17}
\begin{eqnarray}
\Xi(t)\equiv\frac{4}{3t}\Big{\{}1-\ln t+\Big{(}\frac{5}{2}-\frac{3}{2}\ln
t\Big{)}t+\frac{(1+t)^{2}}{t}[L_{2}(-t)+\ln t\ln(1+t)]\Big{\}}\label{Xi}\;,
\label{eq:46}\end{eqnarray}
where $L_{2}(x)$ is the dilogarithmic function.
\begin{eqnarray}
L_{2}(x)=-\int_{0}^{x}dy\frac{\ln(1-y)}{y}\;.
\label{eq:47}\end{eqnarray}
Though we define $\omP(t)$ separately in the IR and UV
domains, the two regions are related by the conformal symmetry
$t\leftrightarrow\frac{1}{t}$.  

The Adler-D function, related to $\Pi(Q^{2})$ through 
Eq.(\ref{eq:20}), will have the one-chain skeleton expansion term 
with characteristic function $\omD(t)$,
\begin{eqnarray}
\MDlpt(Q^{2})=\int_{0}^{\infty}{dt}\;\omD(t)a(t Q^{2})\;.\label{D
spec}
\label{eq:48}\end{eqnarray}
$\omD(t)$ is obtained from $\omP(t)$ by performing the differentiation of Eq.(\ref{eq:20})
 on Eq.(\ref{eq:44}) and then performing integration by
parts on the resulting expression. 
\begin{eqnarray}
\MDlpt(Q^{2})&=&-\frac{3}{4}Q^{2}\frac{d}{dQ^{2}}\int_{0}^{\infty}dt\,\omP(t)t \Bigg{(}\frac{a(t Q^{2})}{t}\Bigg{)}\non\\
&=&+\frac{3}{2b}Q^{2}\frac{d}{dQ^{2}}\int_{0}^{\infty}dt\frac{d}{dt}\Bigg{[}\omP(t)t\Bigg{]}\ln[a(t Q^{2})]\non \\
&=&-\frac{3}{4}\int_{0}^{\infty}dt\Bigg{[}\omP(t)+t\frac{d}{dt}\omP(t)\Bigg{]}a(t Q^{2})\;.
\label{eq:49}\end{eqnarray}
The transformation from $\Pi$ to $\MD$ therefore induces a
transformation in $\omP(t)$ of
\begin{eqnarray}
\PIQ&\rightarrow&Q^{2}\frac{d}{dQ^{2}}\PIQ=-\frac{4}{3}\MD(Q^{2})\non \\
\Rightarrow\qquad\omP(t)&\rightarrow&\omP(t)+t\frac{d}{dt}\omP(t)=-\frac{4}{3}\omD(t)\;.\label{trans}
\label{eq:50}\end{eqnarray}
This transformation spoils the conformal symmetry
present in $\omP(t)$. Indeed the expressions for
$\omD(t)$ in the UV and IR regions are slightly more
complicated.
\begin{eqnarray}
\omD^{IR}(t)&=&\frac{8}{3}\Bigg{\{}\Bigg{(}\frac{7}{4}-\ln t\Bigg{)}t+(1+t)\Big{[}L_{2}(-t)+\ln t\ln(1+t)\Big{]}\Bigg{\}}\label{eq:51}\\
\omD^{UV}(t)&=&\frac{8}{3}\Bigg{\{}1+\ln
t+\Bigg{(}\frac{3}{4}+\frac{1}{2}\ln
t\Bigg{)}\frac{1}{t}\non\\
&+&(1+t)\Big{[}L_{2}(-t^{-1})-\ln
t\ln(1+t^{-1})\Big{]}\Bigg{\}}
\label{eq:52}\end{eqnarray}
However, a partial symmetry remains in $\omD(t)$ and this
will be elucidated upon in the following discussion. We
shall now convert the skeleton expansion form into
the Borel representations of Eqs.(\ref{eq:9}) and (\ref{eq:12}) by making a
change of variables.
To achieve this it is necessary to write $\omP(t)$ as an
expansion in powers of $t$. This yields expressions in both the IR and UV
regions comprising an expansion plus an expansion times a
logarithm.
\begin{eqnarray}
\omP^{IR}(t)&=&-\frac{4}{3}\left(\sum_{n=1}^{\infty}\xi_{n}t^{n}+\ln t\sum_{n=2}^{\infty}\xih_{n}t^{n}\right)\;.
\label{eq:53}\end{eqnarray}
The conformal symmetry expressed in Eq.(\ref{eq:45}) means that the UV part can also
be written in terms of the coefficients $\xi_{n}$ and $\xih_{n}$
\begin{eqnarray}
\omP^{UV}(t)&=&-\frac{4}{3}\left(\sum_{n=1}^{\infty}\xi_{n}t^{-n}-\ln t\sum_{n=2}^{\infty}\xih_{n}t^{-n}\right)\;.
\label{eq:54}\end{eqnarray}
From Eq.(\ref{Xi}), $\xi_{n}$ and $\xih_{n}$ are found
to be
\begin{eqnarray}
\xi_{n>1}&=&\frac{4}{3}\frac{(2-6n^{2})(-1)^{n}}{(n-1)^{2}n^{2}(n+1)^{2}},\qquad\qquad\xih_{n>1}=\frac{4}{3}\frac{2(-1)^{n}}{(n-1)n(n+1)}\label{xi}\non \\
\xi_{1}&=&1\hskip56.5mm\xih_{1}=0
\label{eq:55}\end{eqnarray}
Performing the transformation in Eq.(\ref{eq:50}) allows us
to write $\omD(t)$ as a similar expansion
\begin{eqnarray}
\omD^{IR}(t)&=&\sum_{n=1}^{\infty}[\xi_{n}(1+n)+\xih_{n}]t^{n}+\ln t\sum_{n=2}^{\infty}\xih_{n}(n+1)t^{n}\label{eq:56}\\
\omD^{UV}(t)&=&\sum_{n=1}^{\infty}[\xi_{n}(1-n)-\xih_{n}]t^{-n}+\ln
t\sum_{n=2}^{\infty}\xih_{n}(n-1)t^{-n}\label{omegaDexp}
\label{eq:57}\end{eqnarray}

Using the expansions of Eqs.(\ref{eq:56}) and (\ref{eq:57}) we can now represent
$\MDlpt(Q^{2})$ in terms of a Borel integral. We take $\MDlpt(Q^{2})$
expressed in terms of $\omD(t)$  and then split
the integral into IR and UV regions
\begin{eqnarray}
\MDlpt(Q^{2})&=&\int_{0}^{\infty}dt\,\omD(t)a(t Q^{2})\non \\
&=&\sum_{k=0}^{\infty}a(Q^{2})\int_{0}^{1}dt\,\omD^{IR}(t)\Bigg{(}-\frac{ba(Q^{2})}{2}\ln t\Bigg{)}^{k}\non \\
&+&\sum_{k=0}^{\infty}a(Q^{2})\int_{1}^{\infty}dt\,\omD^{UV}(t)\Bigg{(}-\frac{ba(Q^{2})}{2}\ln t\Bigg{)}^{k}\non \\
&=&a(Q^{2}) \sum_{k=0}^{\infty}\Bigg{(}-\frac{ba(Q^{2})}{2}\Bigg{)}^{k}\non \\
&&\Bigg{[}\int_{0}^{1}dt\Bigg{(}\sum_{n=1}^{\infty}[\xi_{n}(1+n)+\xih_{n}](t)^{n}+\ln t\sum_{n=2}^{\infty}\xih_{n}(n+1)(t)^{n}\Bigg{)}(\ln t)^{k}\non\\
&+&\int_{1}^{\infty}dt\Bigg{(}\sum_{n=1}^{\infty}[\xi_{n}(1-n)-\xih_{n}](t)^{-n}+\ln
t\sum_{n=2}^{\infty}\xih_{n}(n-1)(t)^{-n}\Bigg{)}(\ln t)^{k}\Bigg{]}
\label{eq:58}\end{eqnarray}
Where we have used
\begin{eqnarray}
a(xy)=a(y)\sum_{k=0}^{\infty}\Bigg{(}-\frac{ba(y)}{2}\ln x\Bigg{)}^{k}\;.
\label{eq:59}\end{eqnarray}

We note that $[\xi_{n}(1-n)-\xih_{n}]=0$ for $n=1$, which
allows us to omit this term from the above sum. This expression
may be transformed into a Borel integral of the form of Eq.(\ref{eq:9}) by changes of
variables and integration by parts. We use the change of variables
$z=-a(Q^2)(n+1)\ln t$ and $z=a(Q^2)(n-1)\ln t$ for IR and
UV parts, respectively. Integration by parts is necessary for
the integrals with an extra $\ln t$ term. For $Q^2>{\Lambda}^{2}$, $a(Q^2)>0$, we then obtain the standard
Borel representation, of Eq.(\ref{eq:9})
\begin{eqnarray}
\MDlpt(Q^2)&=&\int_{0}^{\infty}dz\,e^{\frac{-z}{a(Q^2)}}\Bigg{[}\sum_{n=1}^{\infty}\frac{[\xi_{n}(1+n)+\xih_{n}]}{n+1}\frac{1}{1-\frac{bz}{2(n+1)}}-\sum_{n=2}^{\infty}\frac{\xih_{n}(n+1)}{(n+1)^{2}}\frac{1}{\Big{(}1-\frac{bz}{2(n+1)}\Big{)}^{2}}\Bigg{]}\non\\
&+&\!\!\!\!\!\int_{0}^{\infty}dz\,e^{\frac{-z}{a(Q^{2})}}\Bigg{[}\sum_{n=2}^{\infty}\frac{[\xi_{n}(1-n)-\xih_{n}]}{n-1}\frac{1}{1+\frac{bz}{2(n-1)}}+\sum_{n=2}^{\infty}\frac{\xih_{n}(n-1)}{(n-1)^{2}}\frac{1}{\Big{(}1+\frac{bz}{2(n-1)}\Big{)}^{2}}\Bigg{]}\;,\non\\\label{eq:60}
\end{eqnarray}
and for $Q^2<{\Lambda}^{2}$, $a(Q^2)<0$, we obtain the modified Borel representation of Eq.(\ref{eq:12}), in which the upper limit in $z$ is $-\infty$.
Having obtained the Borel transform we can now make contact with
Eq.(\ref{eq:25}) and this allows us to make the
identifications
\begin{eqnarray}
\qquad\qquad\qquad\qquad\frac{\xi_{n}(1+n)+\xih_{n}}{n+1}=-B_{1}(n+1)z_{n+1}\qquad\qquad\qquad\qquad  n\geq1\label{eq:61}\\
\qquad\qquad\qquad\qquad\frac{\xi_{n}(1-n)-\xih_{n}}{n-1}=A_{1}(n-1)z_{n-1}\,\,\,\,\qquad\qquad\qquad\qquad
n\geq2\label{singlePoles}
\label{eq:62}\end{eqnarray}
for the single pole residues and
\begin{eqnarray}
\,\,\qquad\qquad\qquad-\frac{\xih_{n}(n+1)}{(n+1)^{2}}&=&B_{0}(n+1)+B_{1}(n+1)z_{n+1}\qquad\qquad\quad
n\geq2\label{eq:63}\\
\,\,\qquad\qquad\qquad\frac{\xih_{n}(n-1)}{(n-1)^{2}}&=&A_{0}(n-1)-A_{1}(n-1)z_{n-1}\qquad\qquad\quad
n\geq2
\label{eq:64}\end{eqnarray}
for the double pole residues. Substituting the form of $\xi_{n}$ and $\xih_{n}$ given
by Eq.(\ref{eq:55}), and comparison with Eq.(\ref{eq:26}), verifies the above
equations. 

Equations.~(\ref{eq:61}) - (\ref{eq:64}) can be used to
rewrite the  $\omD^{IR}(t)$ and $\omD^{UV}$ expansions of Eqs.(\ref{eq:56})
and (\ref{eq:57}) in terms of the
${A}_{0}(n)$, ${A}_{1}(n)$, and ${B}_{0}(n)$, ${B}_{1}(n)$ renormalon residues. One finds
\begin{eqnarray}
\!\!\omD^{IR}(t)\!\!&=&\!\!\frac{b}{2}\sum_{n=1}^{\infty}-{z}_{n+1}^{2}{B}_{1}(n+1){t}^{n}-{\ln t}\sum_{n=2}^{\infty}
{(n+1)}^{2}[{B}_{0}(n+1)+{z}_{n+1}{B}_{1}(n+1)]{t}^{n}\;\; \label{eq:65}\\
\!\!\omD^{UV}(t)\!\!&=&\!\!\frac{b}{2}\sum_{n=1}^{\infty}{z}_{n-1}^{2}{A}_{1}(n-1){t}^{-n}+{\ln t}\sum_{n=2}^{\infty}{(n-1)}^{2}[{A}_{0}(n-1)-
{z}_{n-1}{A}_{1}(n-1)]{t}^{-n}.\;\;
\label{eq:66}\end{eqnarray}
The discontinuity at $t=1$ is then found to be
\begin{equation}
\omD^{UV}(1)-\omD^{IR}(1)=\frac{b}{2}\sum_{n=1}^{\infty}{z}_{n}^{2}[{A}_{1}(n)+{B}_{1}(n)]\;,
\label{eq:67}\end{equation}
which vanishes using Eq.(\ref{eq:40}). In the language of $\xi_{n}$ and
$\xih_{n}$ coefficients, Eq.(\ref{eq:67}) is equivalent to
\begin{eqnarray}
-2\sum_{n=1}^{\infty}\left(n\xi_{n}+\xih_{n}\right)=0\;.
\label{Eq:67a}
\end{eqnarray}
So the relation between UV and IR renormalon residues of
Eq.(\ref{eq:39}), which guarantees finiteness at $Q^2={\Lambda}^{2}$, ensures that the characteristic
function ${\omD}(t)$ is continuous at $t=1$. 

For the first derivative at $t=1$ one finds the
discontinuity 
\begin{eqnarray}
\frac{d\omD^{IR}}{dt}\Bigg{|}_{t=1}-\frac{d\omD^{UV}}{dt}\Bigg{|}_{t=1}&=&b\sum_{n=1}^{\infty}{z}_{n}^{2}[{A}_{1}(n)+{B}_{1}(n)]
\non \\
&-&\frac{b^2}{4}\sum_{n=1}^{\infty}{z}_{n}^{2}[{A}_{0}(n)+{B}_{0}(n)]+\frac{b^2}{2}\sum_{n=1}^{\infty}{z}_{n}^{3}[{A}_{1}(n)-{B}_{1}(n)]\;,
\label{eq:68}\end{eqnarray}
Equation.~(\ref{eq:40}), which ensures that $\MDlpt({\Lambda}^{2})$ is finite, means that the first line of this
expression vanishes. The second line also vanishes, ensuring continuity of
the first derivative of $\omega_{\mathcal{D}}(t)$. This also ensures that the
${\MD}^{(L)'}_{PT}({\Lambda}^{2})$ is finite (the prime denoting the first
derivative $d/d{\ln}Q$). Indeed, the required relation corresponding to the vanishing
of the coefficient of the potentially 
divergent $\ln a$ term in ${\MD}^{(L)'}_{PT}({\Lambda}^{2})$ is,
\begin{equation}
\sum_{n=1}^{\infty}[2{z}_{n}^{3}({A}_{1}(n)-{B}_{1}(n))-{z}_{n}^{2}({A}_{0}(n)+{B}_{0}(n)]=0\;.
\label{eq:69}\end{equation}
So finiteness of the first derivative of $\mathcal{D}^{(L)}(\Lambda)$
at $Q=\Lambda$, corresponds to continuity of the first derivative of $\omega(t)$ at
$t=1$. Furthermore, Eq.(\ref{eq:69}) written in terms of $\xi_{n}$ and
$\xih_{n}$ is simply Eq.(\ref{Eq:67a}), with an extra factor of
$-2$. Consequently, the continuity of $\omD(t)$ and its first derivative stem
for a single relation, Eq.~(\ref{Eq:67a}). The second and third derivatives are also continuous at $t=1$, and their discontinuities involve 
additional new structures built from combinations of the ${A}_{0,1}$ and ${B}_{0,1}$. To ensure 
finiteness of ${\MD}^{(L)''}_{PT}(Q^2)$ at $Q^2={\Lambda}^{2}$, 
one requires the relation
\begin{equation}
\sum_{n=1}^{\infty}[3{z}_{n}^{4}({A}_{1}(n)+{B}_{1}(n))-2{z}_{n}^{3}({A}_{0}(n)-{B}_{0}(n))]=0\;.
\label{eq:70}\end{equation}
For finiteness of ${\MD}^{(L)'''}(Q^2)$ at $Q^2={\Lambda}^{2}$ one requires the relation
\begin{equation}
\sum_{n=1}^{\infty}[4{z}_{n}^{5}(A_1(n)-B_1(n))-3{z}_{n}^{4}(A_0(n)+B_0(n))]=0\;.
\label{eq:71}\end{equation}
Eqs.(\ref{eq:70}) and (\ref{eq:71}) are also required in order for the second and third derivatives of $\omD(t)$
to be continuous at $t=1$, furthermore, they can both be derived from the following
relation
\begin{eqnarray}
\sum_{n=1}^{\infty}\left(n^{3}\xi_{n}+3n^{2}\xih_{n}\right)=0\;.
\label{Eq:71a}
\end{eqnarray}
The fourth and higher derivatives of $\omD(t)$ are discontinuous at $t=1$ as noted in \cite{r4}.

\section{Skeleton Expansion and the NP component}
In this section we wish to consider more carefully the compensation of ambiguities between renormalons
and the OPE. The regular OPE is a sum over the contributions of condensates with different mass dimensions.
In the case of the Adler function the dimension four gluon condensate is the leading contribution,
\begin{equation}
{G}_{0}(a(Q^2))=\frac{1}{Q^4}\langle 0|GG|0\rangle{C}_{GG}(a(Q^2))\;,  
\label{eq:72}\end{equation}
where ${C}_{GG}(a(Q^2))$ is the Wilson coefficient. In general the $n^{\rm{th}}$ term in the OPE expansion of Eq.(\ref{eq:3})
will have the coefficient 
\begin{equation}
{\MC}_{n}(a(Q^2))={C}_{n}{[a(Q^2)]}^{{\delta}_{n}}(1+O(a))\;.
\label{eq:73}\end{equation}
The exponent $\delta_n$ corresponding to the anomalous dimension of the condensate operator
concerned. Non-logarithmic UV divergences \cite{r18} lead to an ambiguous imaginary part in the
coefficient so that ${C}_{n}={C}_{n}^{(R)}\pm i{C}_{n}^{(I)}$. If one considers an ${IR}_{n}$
renormalon singularity in the Borel plane to be of the form $K_n/{(1-z/{z}_{n})}^{{\gamma}_{n}}$
then one finds an ambiguous imaginary part arising of the form
\begin{equation} 
\textrm{Im}[{\MD}_{PT}]=\pm{K}_{n}\frac{\pi{z}_{n}^{{\gamma}_{n}}}{\Gamma({\gamma}_{n})}{e}^{-{z}_{n}/a(Q^2)}{a}^{1-{\gamma}_{n}}[1+O(a)]\;.
\label{eq:74}\end{equation}
Here the $\pm$ ambiguity comes from routing the contour above or below the real $z$-axis in the
Borel plane. This is structurally the same as the ambiguous OPE term in Eq.(\ref{eq:73}), and if
${C}_{n}^{(I)}=K_{n}\pi{z}_{n}^{{\gamma}_{n}}/\Gamma({\gamma}_{n})$ and ${\delta}_{n}=1-{\gamma}_{n}$, then the PT Borel and
NP OPE ambiguities can cancel against each other \cite{r19}. Taking a PV of the Borel integral corresponds
to averaging over the $\pm$ possibilities. For ${Q^2}<{\Lambda}^{2}$ the modified expansion of Eq.(\ref{eq:15})
will have an $n^{\rm{th}}$ coefficient of the form
\begin{equation}
{\tilde{\MC}}_{n}(a(Q^2))={\tilde{C}}_{n}{[a(Q^2)]}^{{\tilde{\delta}}_{n}}(1+O(a))\;.
\label{eq:75}\end{equation}
Now the exponent ${\tilde{\delta}}_{n}$ is related to the anomalous dimension of
dimension $6$, four-fermion operators associated with UV renormalons \cite{r20}, IR divergences
associated with these render the imaginary part ambiguous, and ${\tilde{C}}_{n}={\tilde{C}}_{n}^{(R)}\pm{\tilde{C}}^{(I)}_{n}$.
The modified Borel representation of Eq.(\ref{eq:12}) has ambiguities arising from UV renormalons.
Assuming that the ${UV}_{n}$ singularity is of the form ${\tilde{K}}_{n}/{(1+z/{z}_{n})}^{{\tilde{\gamma}}_{n}}$ one finds
\begin{equation}
\textrm{Im} [{\MD}_{PT}]=\pm{\tilde{K}}_{n}\frac{\pi{z}_{n}^{{\tilde{\gamma}}_{n}}}{\Gamma({\tilde{\gamma}}_{n})}{e}^{{z}_{n}/a(Q^2)}{a}^{1-{\tilde{\gamma}}_{n}}
[1+O(a)]\;.
\label{eq:76}\end{equation}
This is structurally the same as the ambiguity in the modified NP expansion coefficient in Eq.(\ref{eq:75}), and
if ${\tilde{C}}_{n}^{(I)}={\tilde{K}}_{n}\pi{z}_{n}^{{\tilde{\gamma}}_{n}}/\Gamma({\tilde{\gamma}}_{n})$ and
${\tilde{\delta}}_{n}=1-{\tilde{\gamma}}_{n}$, the ambiguities can be cancelled.\\  

\begin{figure*}
\begin{tabular}{lr}
\hspace{.2cm}
\includegraphics[angle=270,width=0.45\textwidth]{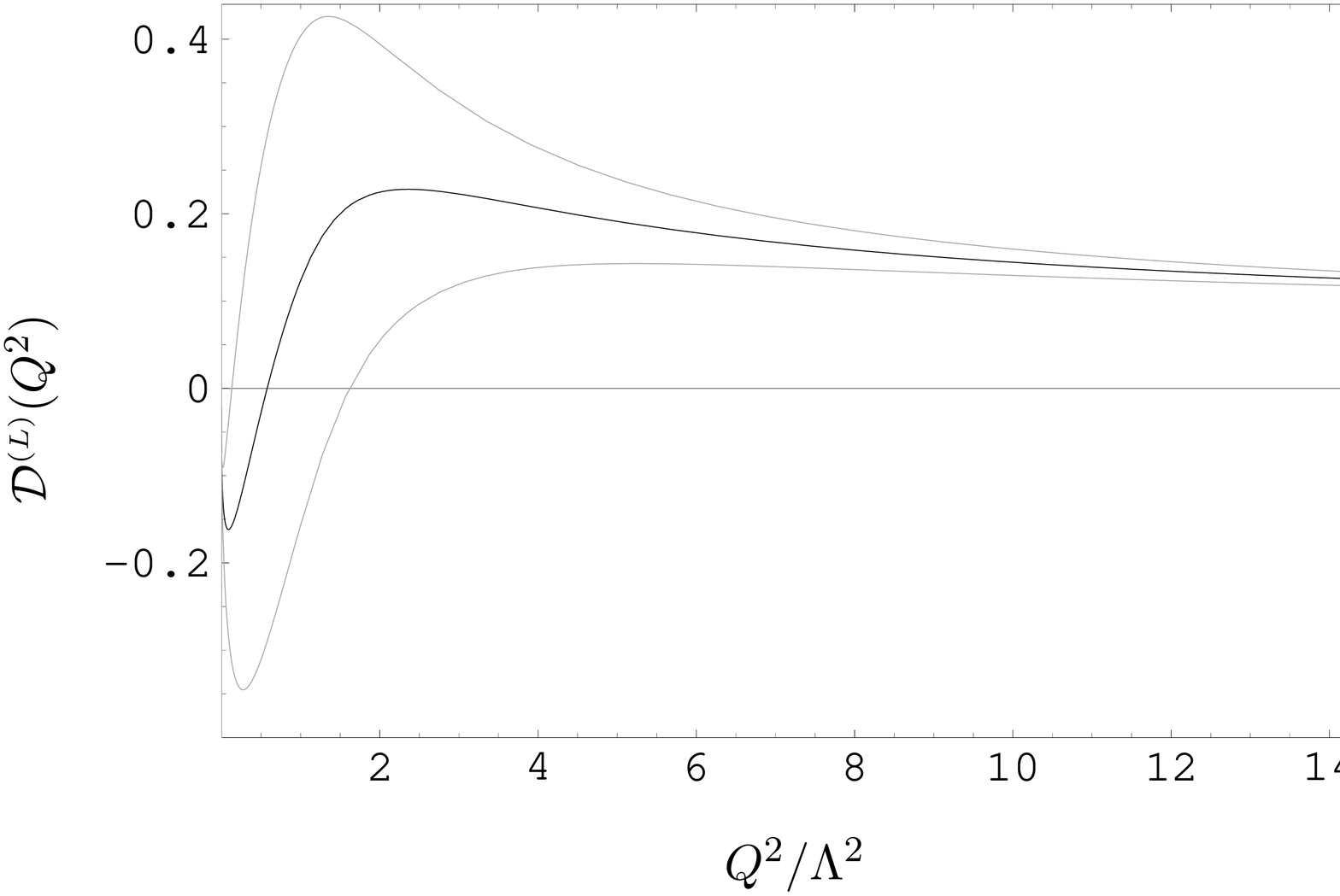}\hspace{.045\textwidth}&\hspace{.045\textwidth}
\includegraphics[angle=270,width=0.45\textwidth]{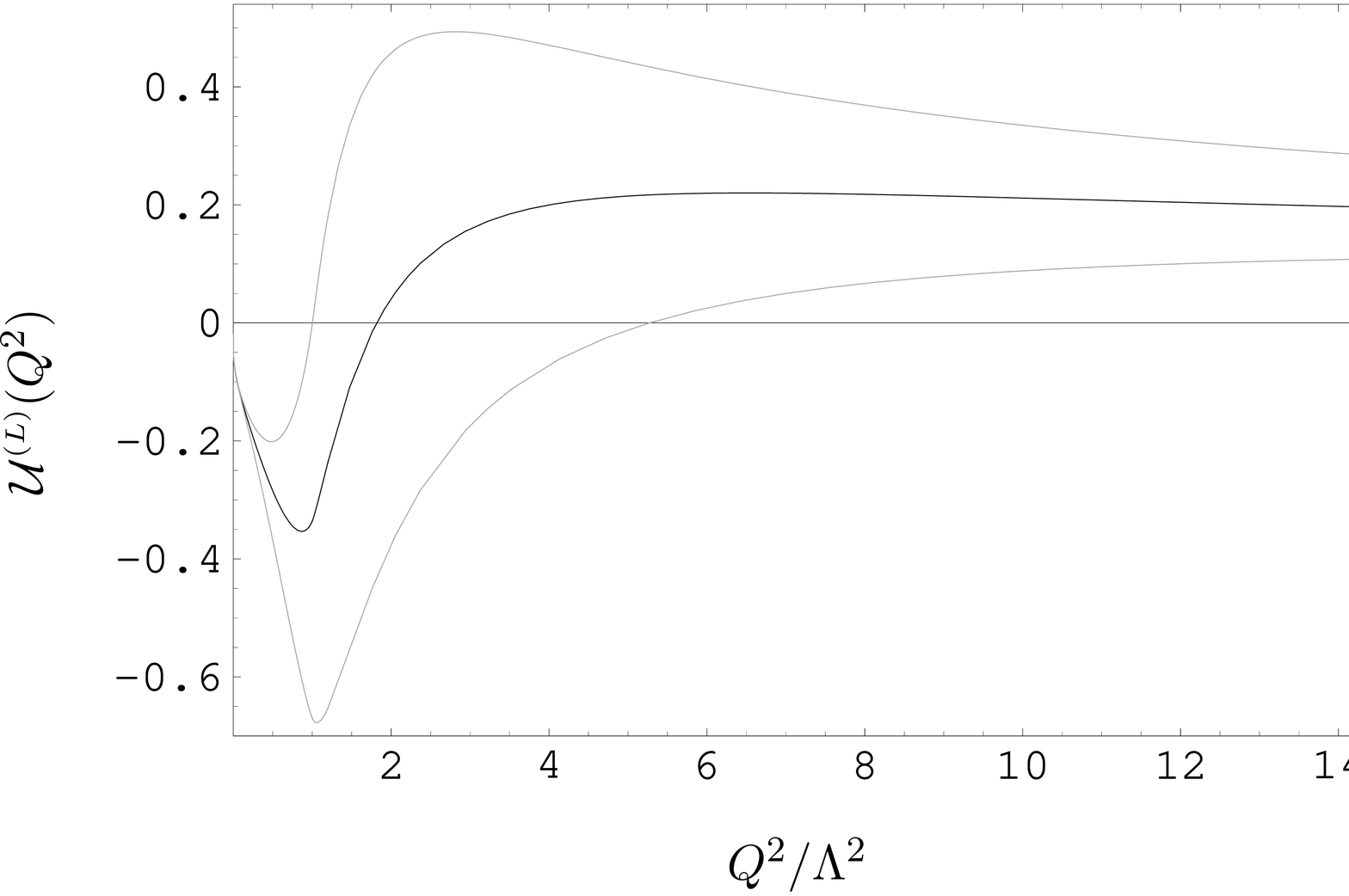}\\
\multicolumn{2}{c}{
\includegraphics[angle=270,width=0.45\textwidth]{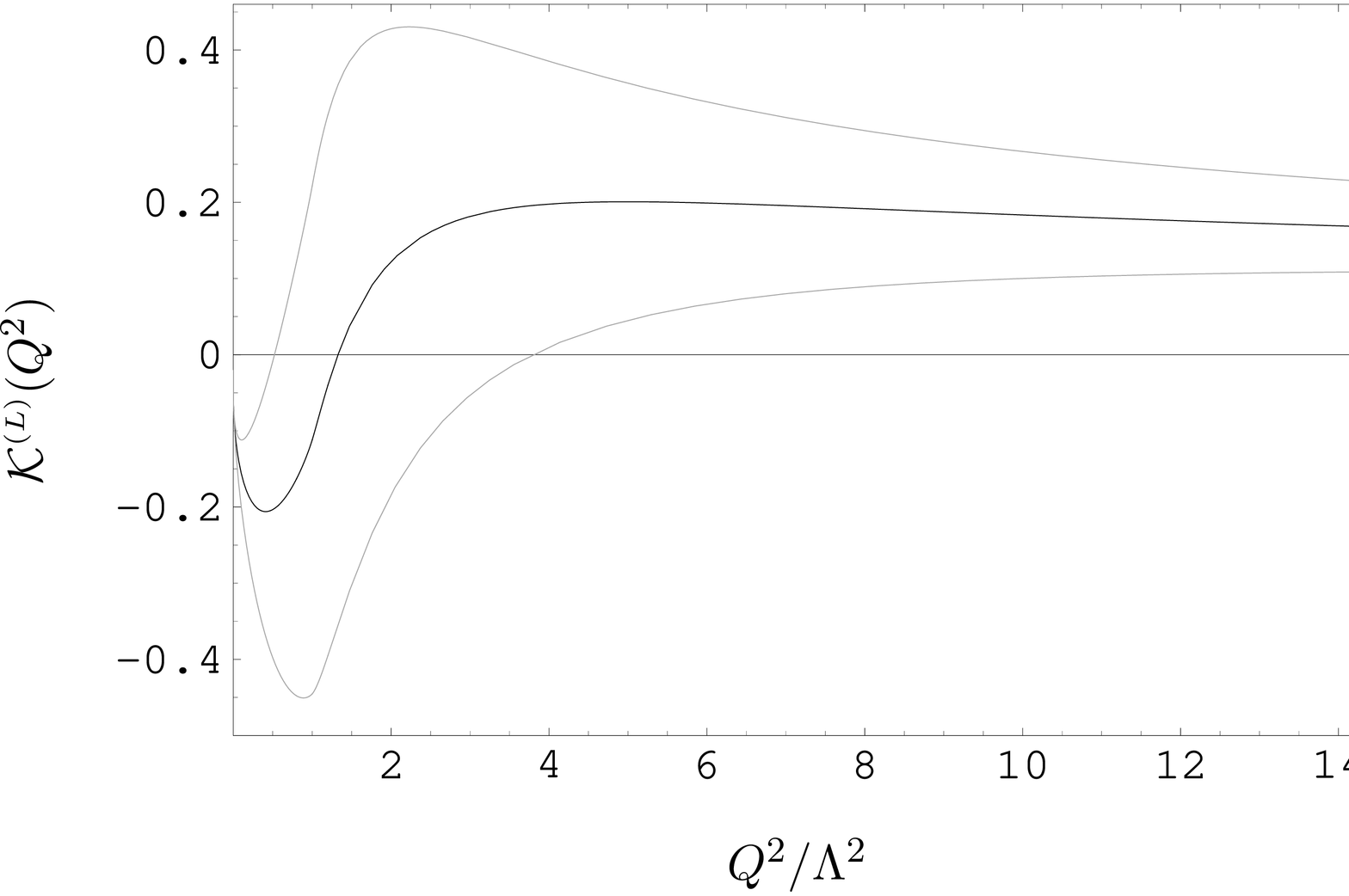}}\\\\
\end{tabular}
\caption{The bold curves corresponding to $\kappa=0$ are the perturbative corrections to the observables in Eqs.(\ref{eq:21})-(\ref{eq:24}), as in Fig. \ref{F:DUK}.
The upper and lower curves correspond to the overall result including NP contributions with the choice $\kappa=1$ and $\kappa=-1$, respectively.}
\label{F:DUKnp}
\end{figure*}

In the one-chain (leading-$b$) approximation the renormalons are single or double poles
corresponding to $\gamma=1$ or $\gamma=2$, and correspondingly the ambiguous imaginary
parts in Eqs.(\ref{eq:74}) and (\ref{eq:76}) contain factors of $a^{1-\gamma}$ which are $1$ or $1/a$, respectively.
For the Adler function $\textrm{Im}[\MDlpt]$
is obtained by making the change $\textrm{Ei}\rightarrow\textrm{Ei}\pm i\pi$ 
in the first line of Eq.(\ref{eq:31}) for $Q^2>{\Lambda}^{2}$, and in the second line for $Q^2<{\Lambda}^{2}$. For
continuity of the $\textrm{Im}$ part at $Q^2={\Lambda}^{2}$ one needs to choose the sign of $i\pi$ oppositely in the two regions.
One then finds for $Q^2>{\Lambda}^{2}$
\begin{eqnarray}
\textrm{Im}[\MDlpt(Q^2)] &=&\pm i\pi\left[
\sum_{n=1}^{\infty}B_{1}(n+1)z_{n+1}^{2}\left(\frac{\Lambda^{2}}{Q^2}\right)^{(n+1)}\right.\non \\
&-&\left.\frac{1}{a(Q^2)}\sum_{n=2}^{\infty}z_{n+1}^{2}[B_{0}(n+1)+z_{n+1}B_{1}(n+1)]\left(\frac{\Lambda^{2}}{Q^2}\right)^{n+1}\right]\;.
\label{eq:77}\end{eqnarray}
Correspondingly, for $Q^2<{\Lambda}^{2}$ one finds
\begin{eqnarray}
\textrm{Im}[\MDlpt(Q^2)]&=&\mp i\pi
\left[\sum_{n=2}^{\infty}A_{1}(n-1)z_{n-1}^{2}\left(\frac{Q^2}{\Lambda^{2}}\right)^{n-1}\right. \non \\
&-&\left.\frac{1}{a(Q^2)}\sum_{n=2}^{\infty}{z}_{n-1}^{2}[{A}_{0}(n-1)-{z}_{n-1}{A}_{1}(n-1)]{\left(\frac{Q^2}{\Lambda^2}\right)}^{n-1}\right]\;.
\label{eq:78}\end{eqnarray}
Comparing these expressions with Eqs.(\ref{eq:65}) and (\ref{eq:66})  one then finds 
that the imaginary part may be written directly in terms of the characteristic function
$\omD(t)$,
\begin{eqnarray}
\textrm{Im}[\MDlpt(Q^2)]&=&\pm \frac{2\pi}{b}\frac{{\Lambda}^{2}}{Q^2}\omD^{IR}\left(\frac{\Lambda^{2}}{Q^2}\right)\;\;\;(Q^2>{\Lambda}^{2}) \non \\
\textrm{Im}[\MDlpt(Q^2)]&=&\pm\frac{2\pi}{b}\frac{{\Lambda}^{2}}{Q^2}\omD^{UV}\left(\frac{\Lambda^{2}}{Q^2}\right)\;\;\;(Q^2<{\Lambda}^{2})
\label{eq:79}\end{eqnarray}
Continuity at $Q^2=\Lambda^2$ then follows from continuity of $\omega(t)$ at $t=1$. 
The ${C}_{n}^{(R)}$, and ${\tilde{C}}_{n}^{(R)}$ coefficients of the OPE and the modified NP
expansion are in principle independent of the imaginary part, but continuity at $Q^2={\Lambda}^{2}$
is dependent upon relations between the ${A}_{0,1}$ and ${B}_{0,1}$ residues,
such as Eqs.(\ref{eq:39}) and (\ref{eq:41}), and the more complicated structures of Eqs.(\ref{eq:69})-(\ref{eq:71}), needed for finiteness of the $Q^2$ derivatives.
Although not strictly necessary for continuity, this continuity follows naturally if we write
\begin{equation}
{\MD}^{(L)}_{NP}(Q^2)=\left(\kappa\pm\frac{2\pi i}{b}\right)\int_{0}^{{\Lambda}^2/Q^2}{dt}\;\left({\omD}(t)+t\frac{d\omD(t)}{dt}\right)\;.
\label{eq:80}\end{equation}
Here $\kappa$ is an undetermined overall real, non-perturbative factor. The $t$ integration here reproduces
the expressions of Eq.(\ref{eq:79}) in the two $Q^2$ regions. If the PT component is PV regulated
one averages over the $\pm$ possibilities, and combining Eq.(\ref{eq:80}) with Eq.(\ref{eq:48}) for $\MDlpt(Q^2)$
one can write down a result for ${\MD}^{(L)}(Q^2)$ for all values of $Q^2$,
\begin{equation}
{\MD}^{(L)}(Q^2)=\int_{0}^{\infty}{dt}\;\left[{\omD}(t)a(tQ^2)+\kappa\left(\omD(t) + t\frac{d\omD (t)}{dt}\right)\theta({\Lambda}^{2}-tQ^2)\right]\;. 
\label{eq:81}\end{equation}
The $Q^2$ evolution is fixed by the non-perturbative constant $\kappa$, and by $\Lambda$. The infrared limit
is ${\MD}^{(L)}(0)=0$, we have already noted that $\MDlpt(0)=0$, the NP component also
freezes to zero since on integrating the second term one finds an IR limit of $\omD^{IR}(1)-\omD^{UV}(1)=0$,
from continuity of the characteristic function at $t=1$. The same expression holds for the other Euclidean
observables ${\MK}^{(L)}_{PT}(Q^2)$ and ${\MU}^{(L)}_{PT}(Q^2)$ on replacing ${\omD}(t)$ by ${\omK}(t)$ and
${\omU}(t)$, respectively. We plot in Fig. \ref{F:DUKnp} the overall result for ${\MD}^{(L)}(Q^2)$,
${\MK}^{(L)}(Q^2)$ and ${\MU}^{(L)}(Q^2)$ for the choices $\kappa=0$, $\kappa=1$ and $\kappa=-1$. For the DIS
sum rules ${\omK}(t)$, ${\omU}(t)$ and their first derivatives are continuous at $t=1$. In the case of $\MU^{(L)}_{NP}(Q^2)$
there are a total of three non-perturbative terms, and hence the continuity of the characteristic function
and its first derivative fixes the form of the function up to an overall constant factor. Thus Eq.~(\ref{eq:81}) does
indeed hold for $\MU^{(L)}(Q^2)$ without conjecturing the form of Eq.~(\ref{eq:80}). 

\section{Infrared freezing behaviour of ${R}_{{e}^{+}{e}^{-}}$}
We turn in this section to a consideration of freezing behaviour of the Minkowskian
quantity ${R}_{{e}^{+}{e}^{-}}$ which was discussed in Ref.\cite{r2}. This treatment was
criticised in Ref.\cite{r3}, which argued that in fact there is an unphysical divergence
in the infrared limit. We wish to address these criticisms. ${R}_{{e}^{+}{e}^{-}}(s)$
will be defined by Eq.(\ref{eq:21}) with the perturbative corrections $\MD(Q^2)$ replaced by
${\MR}(s)$, $\sqrt{s}$ here is the ${e}^{+}{e}^{-}$ c.m. energy. ${\MR}(s)$ is
related to ${\MD}(-s)$ by analytical continuation from Euclidean to Minkowskian. One
may write the dispersion relation
\begin{equation}
{\MR}(s)=\frac{1}{2{\pi}i}{\int_{-s-i\epsilon}^{-s+i\epsilon}}{dt}\frac{{\MD}(t)}{t}\;.
\label{eq:82}\end{equation}
If ${\MD}(t)$ is represented by a Borel representation as in Eq.(\ref{eq:9}) one arrives at
\begin{equation}
{\MR}^{(L)}_{PT}(s)={\int_{0}^{\infty}}{dz}\,{e}^{{-z}/a(s)}\frac{{\sin}({\pi}bz/2)}{{\pi}bz/2}B[\MDlpt](z)\;.
\label{eq:83}\end{equation}
There is now  an extra oscillatory factor of $\sin(\pi{b}z/2)/(\pi bz/2)$ arising from the analytical continuation.
In consequence each {\it individual} IR or UV renormalon contribution at $Q^2={\Lambda}^{2}$ will be finite, and the cancellation
of Eq.(\ref{eq:40}) is not required. One can also analytically continue the one-chain skeleton expansion result
for $\MDlpt(Q^2)$ to obtain
\begin{equation}
{\MR}^{(L)}_{PT}(s)=\frac{2}{\pi{b}}\int_{0}^{\infty}{dt}\;{\omD}(t)\arctan\left(\frac{\pi{b}a(ts)}{2}\right)\;. 
\label{eq:84}\end{equation}
Here the principal branch of $\arctan$ is assumed so it lies in the interval $[-\pi/2,+\pi/2]$, and $\arctan(0)=0$.
This form is equivalent to the Borel representation of Eq.(\ref{eq:83}) for $s>{\Lambda}^{2}$, and to the modified Borel
representation for ${s}<{\Lambda}^{2}$. Notice that the choice of principal branch is crucial if the PV Borel sum is to be
continuous at $s={\Lambda}^{2}$ . The result freezes to the
IR limit ${\MR}^{(L)}_{PT}(0)=0$, since $\arctan(0)=0$ on the principal branch. This freezing limit differs
from that found in the APT approach \cite{r8A}, where a freezing to an IR limit of $2/b$ occurs. This freezing limit was also erroneously
claimed in Ref.~\cite{r2}, but then the PV Borel sum is discontinuous. 
In Ref.[3] 
unphysical singularities in the region $-{\Lambda}^{2}<s<0$ lead to extra terms and they find
\begin{equation}
{\MR}^{(L)}_{PT}(s)=\frac{2}{\pi b}\int_{0}^{\infty}{dt}\;\omega_{\MD}(t)\arctan\left(\frac{\pi{b}a(ts)}{2}\right)
+\frac{2}{b}\int_{0}^{{\Lambda}^{2}/s}{dt}\;{\omD}(t)+\frac{2}{b}\int_{-{\Lambda}^{2}/s}^{0}{dt}\;\omD^{IR}(t)\;.
\label{eq:85}\end{equation}
These extra terms may be treated as contributions to ${\MR}^{(L)}_{NP}(Q^2)$. The final term leads to an infrared divergence as $s\rightarrow{0}$, and has 
an expansion of the same form as the OPE. Notice, however, that the Minkowskian OPE for ${\MR}(Q^2)$ is pathological and
contains delta-functions $\delta(s)$ and their derivatives \cite{r21}. It is only when a smearing procedure in $Q^2$ is used \cite{r22} that it makes sense.    
In contrast for Euclidean quantities the regular OPE is potentially well-defined, and no smearing is required.

We will now consider the evaluation of the PV Borel integral for
${\MR}^{(L)}_{PT}$, and correct the erroneous statements made in Ref.\cite{r2} noted above.
This can be expressed in terms of generalized
exponential integral functions $\textrm{Ei}(n,w)$, defined for $\textrm{Re}\;w>0$ by
\begin{equation}
\textrm{Ei}(n,w)=\int_{1}^{\infty}{dt}\frac{{e}^{-wt}}{t^n}\;.
\label{eq:86}\end{equation}
One also needs the integral
\begin{equation}
\int_{0}^{\infty}{dz}\,{e}^{-z/a}\frac{{\sin}(\pi{b}z/2)}{z}={\arctan}\left(\frac{{\pi}ba}{2}\right)\;.
\label{eq:87}\end{equation}
Here the principal branch of the $\arctan$ is again assumed.
Care needs to be taken when $\textrm{Re}\;w<0$. With the standard continuation 
one arrives at a function analytic everywhere in the cut complex $w$-plane, except at $w=0$
, and with a branch cut running along the negative real semi-axis. Explicitly \cite{r23}
\begin{equation}
\textrm{Ei}(n,w)=\frac{{(-w)}^{n-1}}{(n-1)!}\left[-{\ln}w-{\gamma}_{E}+\sum_{m=1}^{n-1}\frac{1}{m}\right]
-\sum_{\scriptstyle{m}={0}\atop\scriptstyle{m}\neq{n-1}}\frac{{(-w)}^{m}}{(m-n+1)m!}\;.
\label{eq:88}\end{equation}
The $\ln w$ contributes the branch cut along the negative real semi-axis. To obtain the PV
of the Borel integral one needs to compensate for the discontinuity across the branch cut
and make the replacement $\textrm{Ei}(n,w)\rightarrow \textrm{Ei}(n,w)+i\pi sign(\textrm{Im}\;w)$. One
then finds that the IR renormalon contributions to ${\MR}^{(L)}_{PT}(s)$ can be written
in terms of the functions
\begin{eqnarray}
{\phi}_{-}(p,q)&{\equiv}&{e}^{-{z}_{q}/{a}(s)}{(-1)}^{q}\textrm{Im}[Ei(p,-{z}_{q}/a(s)-i{\pi}b{z}_{q}/2)]
\non \\
&-&\frac{{e}^{-{z}_{q}/{a}(s)}{(-1)}^{q}{z}_{q}^{p-1}}{(p-1)!}{\pi}\textrm{Re}[{(({z}_{q}/{a}(s))+i{\pi}b/2)}^{p-1}]\;\theta(s-{\Lambda}^{2})\,.
\label{eq:89}\end{eqnarray}
The UV renormalon contributions can be written in terms of the functions
\begin{eqnarray}
{\phi}_{+}(p,q)&{\equiv}&{e}^{{z}_{q}/a(s)}{(-1)}^{q}\textrm{Im}[Ei(p,{z}_{q}/a(s)+i{\pi}b{z}_{q}/2)]\;.
\non \\
&+&\frac{{e}^{{z}_{q}/{a}(s)}{(-1)}^{q}{z}_{q}^{p-1}}{(p-1)!}{\pi}\textrm{Re}[{((-{z}_{q}/{a}(s))-i{\pi}b/2)}^{p-1}]\,\theta({\Lambda}^{2}-s)\;.
\label{eq:90}\end{eqnarray}
The PV regulated ${\MR}^{(L)}_{PT}(s)$ is then given for {\it all} values of $s$ by
\begin{eqnarray}
{\MR}^{(L)}_{PT}(s)&=&{\MR}^{(L)}_{PT}(s){|}_{UV}+{\MR}^{(L)}_{PT}(s){|}_{IR}
\non \\
&=&\frac{2}{{\pi}b}{\arctan}\left(\frac{{\pi}b{a(s)}}{2}\right)+\frac{2}{{\pi}b}\sum_{j=1}^{\infty}\left({A_0}(j){\phi}_{+}(1,j)
+({A_0}(j)-{A_1}(j){z}_{j}){\phi}_{+}(2,j)\right)
\non \\
&+&\frac{2{B_0}(2)}{{\pi}b}{\phi}_{-}(1,2)+\frac{2}{{\pi}b}\sum_{j=3}^{\infty}\left({B_0}(j){\phi}_{-}(1,j)
+({B_0}(j)+{B_1}(j){z}_{j}){\phi}_{-}(2,j)\right)\;.
\label{eq:91}\end{eqnarray}
Note that the presence of the $\theta$-functions is crucial in
Eqs.(\ref{eq:89}) and (\ref{eq:90}). The terms they multiply are the
extra contributions necessary to obtain the PV when $\textrm{Re}\;w<0$. For $s>{\Lambda}^2$ the second contribution is
required for the IR renormalon contribution, but for $s<{\Lambda}^{2}$ it must be switched off, otherwise
the Borel integral will not be correctly evaluated. With $a(s)<0$ for $s<{\Lambda}^{2}$, $\textrm{Re}\;w<0$ occurs for
the UV renormalon contributions and the extra term must be switched on to obtain a PV regulation of the
UV component. Leaving out the ${\theta}$-function in Eq.(90) would cause an unphysical divergence in the
infrared, and leaving it out in Eq.(91) would cause asymptotic freedom to fail in the ultraviolet.
If the PV is correctly evaluated with $\arctan$ remaining on the principal branch
for ${s}^{2}<{\Lambda}^{2}$ then one obtains ${\MR}^{(L)}_{PT}(0)=(2/\pi{b})\arctan(0)=0$.  Notice that at
first sight the PV result appears to be discontinuous at $s={\Lambda}^{2}$, as the ${\theta}$-function
contributions switch over. However the discontinuity is given by the ${\phi}_{\pm}(1,j)$ terms, and one
finds, upon summing them, a discontinuity 
\begin{equation}
\frac{2}{{b}}\sum_{j=1}^{\infty}[{B}_{0}(j){(-1)}^{j}+{A}_{0}(j){(-1)}^{j}]=\frac{2}{{b}}{B}_{0}(2)=\frac{2}{{b}}\;.  
\label{eq:92}\end{equation}
Here the relation of Eq.(\ref{eq:41}) ensures pairwise cancellations of terms, and ${B}_{0}(2)=1$ is left over. If we remain
on the principal branch, however, the $\arctan$ term has an equal discontinuity of opposite sign, since
$(2/\pi{b})\arctan({+\infty})=1/b$, whereas $(2/\pi{b})\arctan({-\infty})=-1/b$, and overall there is continuity at $s={\Lambda}^{2}$.  
In Ref.\cite{r2} it was wrongly claimed that the PV result is discontinuous at $s={\Lambda}^{2}$, and instead it
was suggested to use a regulation where one throws away the second terms in Eqs.(90) and (91). These terms 
are of the form  ${({\Lambda}^{2}/s)}^{q}$, and ${(s/{\Lambda}^{2})}^{q}$, respectively,
and so they can simply be absorbed into the regular OPE and its modified form.\\    

We finally discuss the ambiguous $\textrm{Im}[{\MR}^{(L)}_{PT}(s)]$. This may be straightforwardly evaluated as
\begin{eqnarray}
\textrm{Im}[{\MR}^{(L)}_{PT}(s)]&=&\pm i\pi\sum_{n=1}^{\infty}[{B}_{0}(n)+{B}_{1}(n){z}_{n}]{z}_{n}{(-1)}^{n}{\left(\frac{\Lambda^2}{Q^2}\right)}^{n}\;\;\;(Q^2>\Lambda^2)
\non \\
\textrm{Im}[{\MR}^{(l)}_{PT}(s)]&=&\mp i\pi\sum_{n=1}^{\infty}[{A}_{0}(n)-{A}_{1}(n){z}_{n}]{z}_{n}{(-1)}^{n}{\left(\frac{\Lambda^2}{Q^2}\right)}^{n}\;\;\;(Q^2<\Lambda^2)\;.  
\label{eq:93}\end{eqnarray}
If one defines $\omD(t)\equiv\omD^{(1)}(t)+\ln t\; \omD^{(2)}(t)$, the split being into the single and double pole renormalon
contributions, then comparing with Eqs.(\ref{eq:65}) and (\ref{eq:66}) one finds
\begin{eqnarray}
\textrm{Im}[{\MR}^{(L)}_{PT}(s)]&=&\pm\frac{2\pi}{b}\frac{\Lambda^2}{s}\omD^{(2)IR}\left(\frac{-\Lambda^2}{s}\right)\;\;\;(s>\Lambda^2)
\non \\
\textrm{Im}[{\MR}^{(L)}_{PT}(s)]&=&\pm\frac{2\pi}{b}\frac{\Lambda^2}{s}\omD^{(2)UV}\left(\frac{-\Lambda^2}{s}\right)\;\;\;(s<\Lambda^2)\;.
\label{eq:94}\end{eqnarray}
Notice that only the double poles contribute since the $\sin(\pi bz/2)/(\pi bz/2)$ analytical continuation term in Eq.~(\ref{eq:82}) contains
zeros at $z=\pm{z}_{n}$ which nullify the single pole contributions.
Whilst the characteristic function $\omD(t)$ is continuous at $t=1$, the $\omD^{(2)}(t)$ function is {\it discontinuous} at $t=-1$.
The discontinuity is $\pm{2}/b$ and arises from the same sum in Eq.(\ref{eq:92}) which gives an apparent discontinuity in the PV ${\MR}^{(L)}_{PT}(s)$
component, although in the PT case this is cancelled by the $\arctan$ term. Thus defined in this way $\textrm{Im}[{\MR}^{(L)}_{PT}(s)]$ is
discontinuous at $s=\Lambda^2$. It would seem that the proper way to proceed is rather to use the dispersion relation of Eq.(\ref{eq:82}) to analytically continue into
the Minkowskian region the expression for ${\MD}^{(L)}{Q^2}$ arrived at in Eq.(\ref{eq:81}). Unfortunately the one-chain skeleton expansion
form for $\MD(Q^2)$ is hard to consistently analytically continue, which was a key motivation for the alternative inverse Mellin representation introduced
in Ref.\cite{r24}. We shall defer further discussion of the more subtle issue
of Minkowskian freezing until a later work. 

\section{Discussion and Conclusions}
We have shown in this paper that in the approximation of the one-chain QCD skeleton
expansion (leading-$b$ approximation), the perturbative corrections to the
parton model result for Euclidean observables undergo a smooth freezing
to an infrared limit of zero. We explicitly studied the Adler function, GLS
sum rule and polarised and
unpolarised Bjorken DIS sum rules as explicit examples, and found that they changed sign in 
the vicinity of $Q^2=\Lambda^2$, and then froze to zero at $Q^2=0$.
Continuity and finiteness at $Q^2={\Lambda}^{2}$
follow from continuity of the characteristic function $\omega(t)$, and its derivatives at
$t=1$. The one-chain term is equivalent to the standard Borel representation of Eq.(\ref{eq:9}) for
$Q^2>{\Lambda}^{2}$, and to the modified Borel representation of Eq.(\ref{eq:12}), previously
proposed in Ref.\cite{r2}, for $Q^2<\Lambda^{2}$. For the Adler function we established
a dictionary between the residues of the IR and UV renormalon singularities, and the
series expansion coefficients of $\omega(t)$. Continuity of $\omD(t)$ and its first three
derivatives at $t=1$ implies relations between the residues of IR and UV renormalon
singularities: Eqs.(\ref{eq:39}), (\ref{eq:41}), (\ref{eq:69}), (\ref{eq:70})
and (\ref{eq:71}). IR renormalons for $Q^2>\Lambda^2$ lie on the contour of
integration in the Borel representation, and similarly UV renormalons lie on the contour of integration
in the modified Borel representation for $Q^2<\Lambda^{2}$. In both cases these singularities lead to an ambiguous imaginary
part in $\MDlpt(Q^2)$, which can be cancelled against an ambiguous imaginary part in the coefficients
of the non-perturbative terms, in the two $Q^2$ regions. The ambiguous imaginary part may be written
directly in terms of the characteristic function, as in Eq.(\ref{eq:79}), and is continuous at $Q^2=\Lambda^{2}$. 
If the real parts of the condensates are
to result in a ${\MD}^{(L)}_{NP}(Q^2)$ which is continuous at $Q^2=\Lambda^2$ this suggests that one should write these in terms
of the characteristic function as well, which led us to conjecture Eq.(\ref{eq:80}) in which there is a real overall non-perturbative
factor $\kappa$ which is undetermined and observable-dependent. All of these properties and results hold
in general for Euclidean observables for which a one-chain result of the form of Eq.(\ref{eq:6}) can be written
down. As pointed out in Ref.\cite{r4} this is not possible for Minkowskian observables such as ${R}_{{e}^{+}{e}^{-}}$,
and in this case the question of freezing is more delicate. There is no characteristic function $\omega_{\MR}(t)$
to underwrite the smooth transition through $s=\Lambda^2$ from UV to IR. 
Indeed the Minkowskian gluon condensate OPE contribution is proportional to ${\delta}^{'}(s)$ \cite{r21}  
, so without a smearing procedure it will
give an apparent infrared divergence as $s\rightarrow{0}$. We corrected some erroneous statements about
the continuity of ${\MR}^{(L)}_{PT}(s)$ at $s=\Lambda^{2}$ made in \cite{r2}.
The issue of Minkowskian freezing is interesting and requires further investigation.\\

An interesting feature of the skeleton expansion representation of Eq.(\ref{eq:6}) concerns the definition of $a(Q^2)$ for 
$Q^2<\Lambda^2$. In this region the result of Eq.(\ref{eq:2}) is not in fact the solution of the RG equation,
but is an analytical continuation of the $Q^2>\Lambda^2$ result. However notice that in evaluating $\MDlpt(Q^2)$
for $Q^2>\Lambda^2$ one is integrating over the region $t<\Lambda^{2}/Q^2$ where the analytically continued 
$a(Q^2)$ is required. Also notice that QED skeleton expansion results can be obtained simply
by interchanging UV and IR renormalons, and the limits $Q^2\rightarrow{\infty}$ and $Q^2\rightarrow{0}$.
The implication is that QCD in the IR energy region is QED-like, and conversely that QED in the UV energy region
beyond the Landau ghost is QCD-like.
Our result of Eq.(\ref{eq:81}) is closely related to successful models for power corrections based on isolating
the IR renormalon ambiguity, such as \cite{r25}, and to the power correction model of \cite{r26,r27}. 
The latter postulates an infrared finite running coupling, and uses a dispersive approach. The infrared
limit of the coupling ${\overline{\alpha}}_{0}$ is a {\it universal} parameter in this picture, whereas
$\kappa$ in our approach is expected to be observable-dependent. A more sophisticated discussion of power
corrections to DIS sum rules has recently appeared in \cite{r27a}. 
Notice that continuity at $Q^2={\Lambda}^{2}$
is the key constraint leading us to suggest Eq.(\ref{eq:80}), which is arguably not a model for power corrections but the actual form of
the NP component in the one-chain (leading-$b$) approximation. In future work we intend to report on
fits of $\kappa$ and $\Lambda$ to experimental data on the Adler function (e.g., the analysis of \cite{r28}), and DIS
sum rules. It would also be interesting to compare our results for the $Q^2$-dependence of the polarised and
unpolarised Bjorken sum rules in the light of the relations between them noted in Ref.\cite{r28a}.
Whilst the freezing is straightforward to analyse in the leading-$b$, one-chain approximation, it is
much harder to analyse at the two-chain level, where the anomalous dimensions for the operators will enter.
The IR$\leftrightarrow$UV conformal relations between renormalon residues and condensate coefficients would
appear to be the key ingredient in the freezing picture, and there is hope that they can continue
to hold at higher orders in the QCD skeleton expansion, if indeed such an expansion can be consistently formulated \cite{r12}.
There is clearly much still to investigate.

\section*{Acknowledgements} We would like to thank Irinel Caprini, Jan Fischer, Andrei Kataev and Dimitri Shirkov
for extremely useful discussions, which have helped us to improve on an earlier version of this paper.
P.M.B. gratefully acknowledges the receipt of a PPARC UK studentship.

\end{document}